\documentclass{pasa}

\usepackage{amsmath}

\usepackage{times}
\usepackage{graphicx,latexsym, amssymb, amscd, psfrag}
\usepackage{color}
\usepackage{morefloats,rotating,float}
\usepackage{multirow,array}
\usepackage{adjustbox}
\usepackage{mathtools}
\usepackage{framed, caption}
\usepackage{subfigure}
\usepackage{aas_macros}

\usepackage{hyperref}%  
\hypersetup{breaklinks,backref,colorlinks,citecolor=blue,linkcolor=blue}% urlcolor=orange %,anchorcolor=black
\usepackage{gensymb} % for \degree command

\title{Plane Polarization in Comptonization process : a Monte Carlo study}

\author[Kumar]{Nagendra Kumar \thanks{Email: nagendra.bhu@gmail.com}
  \affil{No. 1006, Santosh M'house, 9th Cross, Divanarapalya, Gokul Post, Bangalore 560054}
  \affil{Department of Physics, Indian Institute of Science, Bangalore 560012, India}  
  }

\newcommand{\beqn}{\begin{equation}}
\newcommand{\eeqn}{\end{equation}}

\jid{PASA}
\doi{10.1017/pas.YYYY.xxx}
\jyear{YYYY}

\begin{document}

\begin{frontmatter}
  \maketitle

\begin{abstract}

  %%%%% modified one (for 250 words) 13-Jul-2023
 High energies emissions observed in X-ray binaries (XRBs), active galactic nuclei (AGNs) are linearly polarized. The prominent mechanism for X-ray is the Comptonization process.
  We revisit the theory for polarization in Compton scattering with unpolarized electrons, and note that the ($k \times k'$)-coordinate (in which, ($k \times k'$) acts as a $z$-axis, here $k$ and $k'$ are incident and scattered photon momentum respectively) is more convenient to describe it.
  Interestingly, for a fixed scattering plane the degree of polarization PD after single scattering for random oriented low-energy unpolarized incident photons is $\sim$0.33.
  At the scattering angle $\theta$ = 0 or $\theta \equiv$ [0,25$\degree$], the modulation curve of $k'$ exhibits the same PD 
  and PA (angle of polarization) of $k$, 
and even the distribution of projection of electric vector of $k'$ ($k'_e$) on perpendicular plane to the $k$ indicates same (so, an essential criteria for detector designing). 
  We compute the polarization state in Comptonization process using Monte Carlo methods with considering a simple spherical corona.
  We obtain the PD of emergent photons as a function of $\theta$-angle (or alternatively, the disk inclination angle $i$) on a meridian plane (i.e., the laws of darkening, formulated by \citealp{Chandrasekhar1946}) after single scattering with unpolarized incident photons.
To explore the energy dependency we consider a general spectral parameter set corresponding to hard and soft states of XRBs, we find that for average scattering no. $\langle N_{sc}\rangle$ $\sim$1.1 the PD is independent of energy and PA $\sim 90\degree$ ($k'_e$ is parallel to the disk plane), and for $\langle N_{sc}\rangle$ $\sim$5 the PD value is maximum for $i=45\degree$.
  We also compare the results qualitatively with observation of IXPE for five sources.

\end{abstract}


\end{frontmatter}

\section{Introduction}
Active galactic nuclei (AGNs), X-ray binaries (XRBs) comprise a system of a compact object and an accretion disk, where the compact object (black hole BH, or neutron star NS) accretes material via an disk.    
The high energy emission (mainly X-ray $<$100keV) is highly variable, and is generated at the inner region of the disk. The different spectral states 
suggest the Comptonization process (i.e., up-scattering of low-energy photons
by hot electron gas) 
for generating the high energy emission.
The spectral features, variability timescales, and the nature of variability over different energy band provide insight into, in general, the radiative process and the geometry of the emission region, hence %these
constrain the existing theoretical model \cite[see for review][]{Done-etal2007, McClintock-Remillard2006}.
Three parameters, the seed photon source temperature $T_b$, Comptonizing medium/ corona temperature $T_e$, and optical depth of medium ($\tau$) or the average scattering number $\langle N_{sc} \rangle$ that experienced by photon inside the corona are mainly determined the Comptonization. The generated spectrum generally degenerates over physically motivated emission region geometries which are differed by, mainly, the location $\&$ geometry of, either the seed photon source, or corona, or both.
The combine constraint due to the spectral and energy dependent variability is not sufficient to lift out these degeneracy concretely \cite[e.g.][references therein]{Kumar-Misra2016b}.
In literature, the widely studied corona geometries are lamp-post corona situated at the rotation axis of the BH,   spherical corona,  an extended corona on top of the disk or other disk-corona geometry differed by shape and size. In addition, a  static vs dynamic corona (meant, the corona has a bulk motion) has been also invoked. %In literature,
For example, the observed  high energy ($>$100keV) power-law tail emission in XRBs has been described by both static and dynamic corona, like energetically coupled disk-corona, or hybrid electron distribution \cite[e.g.,][]{Done-etal2007}, or bulk Comptonization (BMC) with relativistic inflow onto compact object, or BMC with relativistic conical outflow \cite[][references therein]{Kumar2017}.
  There is also an uncertainty over the location of the seed photon source, e.g., in NS XRBs two different types of seed photon source %for Comptonization
have been advocated,  one is boundary layer (Hot-seed), and other one is accretion disk (Cold-seed photon model) \cite[e.g.][]{Lin-etal2007}.

X-ray polarization measurement  provides two different independent parameters, degree of polarization
PD and the angle of polarization PA, thus it will provide  extra
constraints on the existing theoretical models along with parameter $-$ spectra and time variability. 
Many another fields, like particle acceleration physics, the prompt emission of gamma-ray bursts (GRBs), hard X-ray emission from millisecond pulsar, magnetized white dwarf (WD) and neutron stars,  are target of opportunity for X-ray polarimetry \cite[e.g.,][]{Fabiani-2018, Krawczynski-etal2019, Chattopadhyay2021}. In this work, the main focus is the X-ray polarized emission from XRBs. 
In literature, X-ray spectra along with polarization have been computed for different aspects of disk-corona geometry (for XRBs, or AGNs) with or without taking account of general relativistic effect \cite[e.g.,][]{Dovciak-etal2011,Tamborra-etal2018}.
\cite{Li-etal2009} have discussed the X-ray polarized emission from the geometrically
thin disk, and 
commented that the degree of polarization decreases with decreasing disk-inclination angle and the angle of polarization for low energies scattered photon is parallel to the disk plane.
\cite{Schnittman-Krolik2010} have computed the X-ray polarization for the hard/SPL (steep power law) state of black hole XRBs with three different corona geometries,  and found that for photon energies above the disk thermal peak the angle of polarization transits to perpendicular to the disk plane from parallel at low energy while the maximum degree of polarization is obtained at higher energy band ($\sim$100 keV) and high inclination angle, e.g., the maximum PD  $\sim 10 \%$ for wedged corona geometry, $\sim 4 \%$  for clumpy geometry and $\sim 4 \%$ for spherical geometry.
\cite{Beheshtipour-etal2017} have predicted that the polarization fraction and angle depend on the shape and size of corona geometry (e.g., wedge and spherical) for a fixed energy spectrum.

For astrophysical sources, it is expected that the high energy emission generated by the Compton scattering process would be linearly polarized as in most cases the orientation of electron spin is random. 
The linearly polarized X-ray emission has been observed in X-ray bright sources. First source is the Crab nebula, which is measured by \cite{Weisskopf-etal1978}, almost 45 years ago, using the {\it OSO} 8 graphite crystal polarimeters at 2.6 and 5.2 keV (see references therein for other sources \citealp{Weisskopf2018}; and for review \citealp{Lei-etal1997}).
The Crab polarization has been measured %in high energy band
by instruments, {\it INTEGRAL/IBIS} \cite[200-800 keV; e.g.,][]{Forot-etal2008}, {\it INTEGRAL/SPI} \cite[130-440 keV][]{Jourdain-Roques2019}, {\it AstroSat/CZTI} \cite[100-380 keV][]{Vadawale-etal2018}, {\it PoGO +}, a balloon-borne polarimeter, \cite[20-160 keV][]{Chauvin-etal2018}, {\it Hitomi/SGD} \cite[60-160 keV][]{Hitomi-Collaboration2018}, IXPE \cite[2-8 keV][]{Bucciantini-etal2023}, PolarLight \cite[3-4.5 keV][]{Feng-etal2020}.
The linear X-ray polarization of Cygnus X-1 
has been measured by the {\it PoGO +} balloon-borne polarimeter in energy band 19-181 keV \cite[][]{Chauvin-etal2018}, here authors favour the extended spherical corona geometry over the lamp-post corona model for high energies emission \cite[see also for gamma-ray linear polarization of Cygnus X-1 measured by {\it INTEGRAL}][]{Laurent-etal2011, Jourdain-etal2012}.
The linear gamma-ray polarization for many bright gamma-ray burst (GRBs) sources has been detected by {\it AstroSat/CZTI} \cite[e.g.][]{Sharma-etal2020, Chattopadhyay-etal2019, Chand-etal2019}, by {\it INTEGRAL} /SPI \cite{McGlynn-etal2007} /IBIS \cite{Gotz-etal2014}, by {\it POLAR} \cite{Zhang-etal2019}, by other instruments, e.g., {\it GAP} \cite[see in details][]{Chattopadhyay-etal2019}.
Recently, IXPE has measured polarization properties of many XRBs, AGNs, pulsar in 2-8 keV energy band
\cite[]{Weisskopf-etal2022,Rawat-etal2023,Marshall-etal2022,Jayasurya-etal2023,Pal-etal2023, Marinucci-etal2022, Doroshenko-etal2022}, and for few sources the polarization is an energy dependent. %
\cite{Long-etal2022} quantified the polarized emission of Sco X-1 using PolarLight observations in 3$-$8 keV, and noted an energy dependent polarization.
The X-ray polarimetry is mainly based on three techniques diffraction, photoelectric effect, and Compton scattering \cite[see for review for working, and forthcoming dedicated mission]{Fabiani-2018}, 
e.g., {\it POLIX} , a Compton scattering based X-ray polarimetry % mission
and one of instrument of recently launched {\it XPoSat}\footnote{https://www.isro.gov.in/XPoSat.html} 
\cite[][]{Paul-etal2016}. 

In this work, we explore the polarization properties of Comptonized photons. We first revisited the theory of plane/linear polarization in Compton scattering. We noticed that the scattered photon with the scattering angle $\theta$ = 0 (or, $< 25\degree$) exhibits the same polarization properties of incident photon. We obtain the laws of darkening of single scattered unpolarized photons (originally formulated by \cite{Chandrasekhar1946}) by discussing the step by step simple cases. We estimate the energy dependency of polarization for single-/ multi- scattered unpolarized photons with considering a simple spherical geometry, we also compare the results with observations.
In the next section, we revisit the theory of polarization for Compton scattering and in \S \ref{mc:method} we describe the Monte Carlo method for Compton scattering with polarization. %information.
In \S \ref{sec:verif} we compare the MC results with theoretical results for single scattered photon.
\S \ref{sec:modu} presents the modulation curve of single scattered photon in perpendicular plane of fixed incident photon's direction.
\S \ref{sec:Chandra} presents the polarization of emergent single scattered photons from a given meridian plane. In \S \ref{sec:multi} we present the energy dependency of polarization for multi scattering events, and make a comparison with the
observations, followed by our summary and conclusions in \S \ref{sec:sum}. 
% \clearpage

\section{Revisited theory of Polarization in Compton scattering}\label{sec:2}

The Compton scattering with unpolarized electrons generates  linearly or plan polarized scattered photons.
For a polarized electron the scattered photon is mainly circularly
polarized \cite[][]{Tolhoek-1956}. The unpolarized electron means that the electrons spin are pointed isotropically in all directions. In this work, we consider only unpolarized electron for the Compton scattering process, 
The Klein-Nishina differential cross section for the plane polarization for free electron at rest %rest frame
is expressed as \cite[e.g.,][]{McMaster-1961, Akhiezer-Berestetskii1965}

\beqn \label{gen-dif-polz}
\frac{d\sigma}{d\Omega} = \frac{1}{4}r_o^2\left(\frac{k'}{k}\right)^2\left[\frac{k}{k'}+\frac{k'}{k}-2+4 \cos^2\Theta \right]
\eeqn

\begin{figure}
\centering
\begin{tabular}{c}%\hspace{-1.5cm}
  \includegraphics[width=0.34\textwidth]{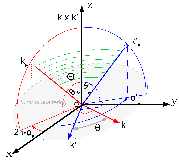}%fig1ff.eps
\end{tabular}\vspace{-0.3cm}
\caption{A schematic diagram for computation in local ($k \times k'$) coordinate where the ($k \times k'$) acts as $z$-axis. In this coordinate the ($x,y$)-plane is a scattering plane, shown by the gray area. The perpendicular plane to $k$ is a red quarter circle (or, plane containing $k_e$ and $(k \times k')$), and the blue quarter circle is a perpendicular plane  of $k'$ (or, plane containing $k'_e$ and $(k \times k')$). The polarization angle is $\theta_e$ and $\theta'_e$ for $k$ and $k'$ respectively and measured with respect to $(k \times k')$. The $\Theta$ is the angle between $k_e$ and $k'_e$, the plane containing $k_e$ and $k'_e$ is shown by the green dotted lines. The direction of $k_e$ and $k'_e$ are $(\theta_e, \phi_e)$ and $(\theta'_e, \phi'_e)$ respectively, thus the scattering angle $\theta$ is $\theta$ = $(\phi_e-\phi'_e)-\pi$.  }
\label{fig:1ang}
\end{figure}

Here, $k=\frac{h\nu}{c}$ is the incident photon momentum, $k'=\frac{h\nu'}{c}$ is the scattered photon momentum, $\nu$ and $\nu'$ are incident and scattered photon frequency, $h$ is the plank constant, $c$ is the speed of light, $\Theta$ is the angle between electric vector of scattered ($k'_e$) and incident ($k_e$) photon, $r_o = \frac{e^2}{mc^2}$ is the classical radius of the electron, e is the elementary charge, m is the mass of the electron, and $d\Omega$ is the differential element of solid angle.
The angle $\Theta$ 
can be determined in terms of angle made by respective electric vectors with ($k$ $\times$ $k'$)-axis (or perpendicular direction to the scattering plane) as, see Figure \ref{fig:1ang},

\beqn \label{cos:pol}
\cos \Theta = \cos\theta_e \cos\theta'_e + \sin\theta_e \sin\theta'_e \cos(\phi_e-\phi'_e)
\eeqn
Here $\theta_e$ and $\theta'_e$ are the $\theta$-angle of electric vector of incident and scattered photon with ($k$ $\times$ $k'$) direction (which acts as a $z$-axis) respectively; and $\phi_e$ and $\phi'_e$ are corresponding $\phi$-angles, which are related to the scattering angle $\theta$ as  $\phi_e - \phi'_e$ $\equiv$ $\pm \pi + \theta$.
The scattered frequency is determined in electron rest frame as
\beqn \label{eq:sctfq}
\frac{\nu'}{\nu} = \frac{1}{1+\frac{h\nu}{m_e c^2} (1-\cos \theta)}
\eeqn

The polarized radiation is uniquely described by four Stokes parameters, $I, Q, U,$ and $V$, with constraint % These parameters are related as
$I = \sqrt{Q^2 + U^2 + V^2}$.
It is defined as $I \equiv I_x + I_y$; $Q \equiv I_x - I_y$; $U \equiv I_{x+45}-I_{y+45}$, and $V$ is a measured of the circular polarization, thus for the present study $V = 0$.
Here, $I_x, I_y$ are the intensity measured along the one of polarized direction, say, along the x-axis and perpendicular to it (or along the y-axis, as here we assume that the photon is travelling along the z-axis); $I_{x+45}, I_{y+45}$ are the intensity measured along the direction which obtains by rotating the x- and y-axis with 45 degree respectively.
The degree of polarization $P$ and the angle of polarization $\chi$ are defined as \cite[e.g.,][]{Lei-etal1997, Bonometto-etal1970}
\beqn \label{p-chi}
P = \frac{\sqrt{Q^2+U^2+V^2}}{I} \quad \ \ \quad \tan (2\chi) = \frac{U}{Q}.\eeqn
For an unpolarized radiation, $P$ = 0, as $I_x = I_y = I_{x+45} = I_{y+45}$. For a partially polarized radiation, and $P = Q/I$, the $P$ varies from -1 to 1, here $P = 1$ is for the completely polarized radiation with electric vector along the $x$-axis and $P = -1$ is for the electric vector along the $y$-axis.

In Compton scattering, to define the Stokes parameters, customary we choose one linear polarization direction is perpendicular to the plane of scattering (or along the
($k \times k'$) direction) and another one is parallel to the scattering plane (i.e., the electric vector of scattered/ incident photon lies on the scattering plane) and the corresponding measured intensity is denoted in terms of differential cross section by  $\sigma_\perp$ ($I_\perp)$ and $\sigma_\parallel$ ($I_\parallel$) respectively. In present notation, the $\theta'_e$ = 0 and $\pi/2$ for $\sigma_\perp$ and $\sigma_\parallel$ respectively. Since, for unpolarized electrons, we have one of Stokes parameter $V = 0$, also in next section we will show that either $U$ = 0 (for unpolarized incident photons) or $U \ll Q$ (for polarized low-energy incident photons). Therefore, in  general, for the  partial polarized photons (a mixture  of polarized and unpolarized photons) 
the degree of polarization for Compton scattering with unpolarized electrons can be written as (\cite[e.g.][]{Lei-etal1997, Dolan-1967})
\beqn \label{deg-polz}
P = \frac{Q}{I} = \frac{I_\perp - I_\parallel}{I_\perp + I_\parallel} = \frac{\sigma_\perp - \sigma_\parallel}{\sigma_\perp + \sigma_\parallel}
\eeqn

The angle of polarization $\chi$ of the scattered photons also measures the angle between two consecutive scattering planes \cite[e.g.,][]{McMaster-1961}.
In other words, the angle between $(k \times k')$ and $(k \times k')_{next}$ is $\chi$. 
Since, the $k'_e$,  $(k \times k')$ and $(k \times k')_{next}$ all are lied in perpendicular plane to $k'$, therefore for next scattering
\beqn \label{chi_th_next} (\theta_e)_{next} = \theta'_e \pm \chi. \eeqn
However  for incident photons, there is no information of previous scattering, only one has $k$ and $k_e$. In computation, for first scattering one has to define the scattering plane freshly, without loss of generality we assume that the angle of polarization ($\chi_{previous}$)  of incident photons is $\theta_e$ with considering $(\theta'_e)_{previous}$ = 0.  
Here the subscript {\it previous} and {\it next} is used for the quantity related to  the previous and next scattering respectively. For clarity, we denote the angle of polarization of the incident photons by $\phi$.

\subsection{Compton scattering of unpolarized photons}
For an unpolarized incident photons %the distribution 
the $\sigma_\perp$ and $\sigma_\parallel$ can be determined by averaging the $\cos^2\Theta$-term of equation (\ref{gen-dif-polz}) over $\theta_e$ using equation (\ref{cos:pol}) for $\theta'_e$ = 0 and $\pi/2$ respectively, and it is epressed as \cite[see, e.g.,][]{McMaster-1961}
\[ d\sigma_\perp^{unpol} = \frac{1}{4}r_o^2\left(\frac{k'}{k}\right)^2\left[\frac{k}{k'}+\frac{k'}{k}\right], \]
\[d\sigma_\parallel^{unpol} = \frac{1}{4}r_o^2\left(\frac{k'}{k}\right)^2\left[\frac{k}{k'}+\frac{k'}{k}-2\sin^2\theta\right], \]
here,  
 $\langle \cos^2\theta_e\rangle$ = $\langle \sin^2\theta_e\rangle$ =0.5; and $\langle \cos\theta_e\rangle$ = $\langle \sin\theta_e\rangle$ =0, as for the unpolarized photons the $\theta_e$ is distributed isotropically.

{\tt The angle of polarization:} Similarly, %, as above,
we estimate the 
$\sigma_{\perp+45}^{unpol}$ and $\sigma_{\parallel+45}^{unpol}$ with having $\theta'_e$ = $\pi/4$ and $3\pi/4$ respectively, its values are $\sigma_{\perp+45}^{unpol} = \sigma_{\parallel+45}^{unpol} = \frac{1}{4}r_o^2\left(\frac{k'}{k}\right)^2\left[\frac{k}{k'}+\frac{k'}{k}-\sin^2\theta\right]$.
Thus the Stokes parameter $U$ is zero, 
which gives $\chi = 0$.
Therefore after single scattering, the scattered photons are polarized along the $(k \times k')$ direction  
or in another words, the plane of polarization of scattered photons is along the perpendicular to
the scattering plane. % for unpolarized incident photons.

{\tt The degree of polarization:} The degree of linear polarization of the scattered photons after single scattering of unpolarized photons is written (using equation \ref{deg-polz}) as  \cite[see, e.g.,][]{McMaster-1961, Matt-etal1996, Lei-etal1997}
\beqn \label{deg-polz-un}
P = \frac{\sin^2\theta}{\frac{k}{k'}+\frac{k'}{k}-\sin^2\theta}.
\eeqn
Here, $P$ = 0, for $\theta$ = 0, and $P = \frac{1}{\frac{k}{k'}+\frac{k'}{k}-1}$ for $\theta$ = 90$\degree$. 
In Thomson limit (precisely defined as $\frac{h\nu}{\gamma \ m_e c^2} \ll 1$, here $\gamma = 1/\sqrt{(1 - \frac{\text v^2}{c^2})}$ is the electrons Lorentz factor, $\text v$ is the speed of electron), one has $\frac{k}{k'} \sim 1$ (see equation \ref{eq:sctfq}), 
thus P = 1 for $\theta = 90\degree$.
Hence, in Thomson regime the single scattered unpolarized (incident) photons at $\theta =90\degree$ are completely polarized in a perpendicular plane to the scattering plane.

{\tt Modulation curve:} The modulation curve is a distribution of the $(\theta'_e)$-angle. 
The differential cross section for unpolarized incident photons can be expressed by using equations (\ref{gen-dif-polz}) and (\ref{cos:pol}) as
\[d\sigma =  \frac{1}{4}r_o^2\left(\frac{k'}{k}\right)^2\left[\frac{k}{k'}+\frac{k'}{k}- 2\sin^2\theta \sin^2\theta'_e\right],\]
here, we consider once again $\langle \cos^2\theta_e\rangle$ = $\langle \sin^2\theta_e\rangle$ =0.5; and $\langle \cos\theta_e\rangle$ = $\langle \sin\theta_e\rangle$ =0. The above expression after rearranging the term can be written as
(using, $\cos 2\theta'_e = 1 -2\sin^2\theta'_e)$ %for the modulation curve
\beqn \label{mod_un} 
d\sigma =\frac{1}{4}r_o^2\left(\frac{k'}{k}\right)^2\left[\frac{k}{k'}+\frac{k'}{k}-\sin^2\theta \right] (1 + P \cos 2\theta'_e)
\eeqn
By comparison to the above expressed modulation curve from the commonly used expression for modulation curve in literature (\citealp[e.g., equation (4.10) of][]{Lei-etal1997}, \citealp[or, equation (2) of][note there, authors have measured the corresponding $\theta'_e$ with respect to the scattering plane]{Chattopadhyay-etal2014}), we again find that the angle of polarization for scattered photons after single scattering of unpolarized incident photons is zero, i.e. the linear polarization is along the perpendicular direction to the scattering plane.

{\tt For polarization-insensitive detector:}
The cross section for a polarization-insensitive detector can be written (by using equations (\ref{gen-dif-polz}) and (\ref{cos:pol}), and now with having additional $\langle \cos^2\theta'_e\rangle$ = $\langle \sin^2\theta'_e\rangle$ =0.5) as
\beqn \label{cross_unpolz}
d\sigma_{ins-detct} = \frac{1}{4}r_o^2\left(\frac{k'}{k}\right)^2\left[\frac{k}{k'}+\frac{k'}{k}-\sin^2\theta\right]
\eeqn
Also, here \begin{footnotesize}$d\sigma_{ins-detct} = (d\sigma_\perp^{unpol} + d\sigma_\parallel^{unpol})/2 = (d\sigma_{\perp+45}^{unpol} + d\sigma_{\parallel+45}^{unpol})/2 $\end{footnotesize}.

\subsection{Compton scattering of polarized photons}\label{sec2:pol}
For a completely polarized incident photons with polarization angle $\phi$ (i.e., $\theta_e = \phi$) the cross section can be obtained by averaging the equation (\ref{gen-dif-polz})  over  $\theta'_e$ \cite[see, e.g.,][and reference therein]{Lei-etal1997}, and it is written as
\beqn \label{eq:cros_pol}
\frac{d\sigma}{d\Omega}^{pol} = \frac{1}{4}r_o^2\left(\frac{k'}{k}\right)^2\left[\frac{k}{k'}+\frac{k'}{k}-2\sin^2\phi \sin^2\theta \right],
\eeqn
here we consider $\langle \cos^2\theta'_e\rangle$ = $\langle \sin^2\theta'_e\rangle$ = 0.5; and $\langle \cos\theta_e\rangle$ = $\langle \sin\theta_e\rangle$ =0. However, it is expected that the distribution of $\theta'_e$ is no longer isotropic but depends on the cross section, equation (\ref{gen-dif-polz}) \cite[see, e.g.,][]{Matt-etal1996}.
Similar to the unpolarized incident photons case, we compute the
$d\sigma_\perp$ and $d\sigma_\parallel$ for polarized incident photons with having
$\theta'_e$ = 0 and $\pi/2$ respectively, which is written as \cite[see, e.g.,][]{McMaster-1961}
\[\begin{aligned}
d\sigma_\perp^{pol} =\frac{1}{4}r_o^2\left(\frac{k'}{k}\right)^2\left[\frac{k}{k'}+\frac{k'}{k}-2+4 \cos^2\phi \right]\\
d\sigma_\parallel^{pol}=\frac{1}{4}r_o^2\left(\frac{k'}{k}\right)^2\left[\frac{k}{k'}+\frac{k'}{k}-2+ 4\sin^2\phi \cos^2\theta \right]
\end{aligned}\]

{\tt The angle of polarization:} The $d\sigma_{\perp+45}$ and $d\sigma_{\parallel+45}$ for polarized incident photons are \begin{footnotesize}$d\sigma_{\perp+45}^{pol} = \frac{1}{4}r_o^2\left(\frac{k'}{k}\right)^2\left[\frac{k}{k'}+\frac{k'}{k}-2+2(\cos^2\phi +\sin^2\phi\cos^2\theta +\sin2\phi \cos\theta) \right]$, and $d\sigma_{\parallel+45}^{pol} = \frac{1}{4}r_o^2\left(\frac{k'}{k}\right)^2\left[\frac{k}{k'}+\frac{k'}{k}-2+2(\cos^2\phi +\sin^2\phi\cos^2\theta -\sin2\phi \cos\theta) \right]$.\end{footnotesize} Here,
$\theta'_e$ = $\pi/4$ and $3\pi/4$ for $d\sigma_{\perp+45}^{pol}$ and $d\sigma_{\parallel+45}^{pol}$ respectively. The Stokes parameters $U$ $\&$ $Q$ are expressed as
\[ U = \frac{1}{4}r_o^2\left(\frac{k'}{k}\right)^2\left[4\sin2\phi \cos\theta) \right] =d\sigma_{\perp+45}^{pol} - d\sigma_{\parallel+45}^{pol} \]
\[ Q = \frac{1}{4}r_o^2\left(\frac{k'}{k}\right)^2\left[4\cos^2\phi-4\sin^2\phi\cos^2\theta\right] =d\sigma_{\perp}^{pol} - d\sigma_{\parallel}^{pol}\]
The angle of polarization can be obtained by using expression (\ref{p-chi}). In practice, the average angle of polarization $\langle \chi\rangle$ is interested, it is written as \cite[e.g.][]{Li-etal2009}
\beqn \label{av_pol}
\tan2\langle \chi\rangle = \frac{\langle U \rangle}{\langle Q \rangle},
\eeqn
here, $\langle U \rangle$ and $\langle Q \rangle$ are averaged of $U$ and $Q$ over angle respectively. In Thomson limit, we find
that the magnitude of $\langle U \rangle$ is almost one order less than the magnitude of $\langle Q \rangle$, i.e., $|\langle U \rangle| << |\langle Q \rangle|$. Thus, 
 $\langle \chi \rangle$ $\sim$ 0 or $\pi/2$ for a positive or negative value of $\langle Q \rangle$ respectively. 

 {\tt The degree of polarization:} 
   On average $|\langle U \rangle| < < |\langle Q \rangle|$, but we notice also $U > Q$ for a  range of $\theta$, e.g., see Figure \ref{fig:uq} of appendix \ref{pl-uq}. Hence we define the PD with considering two extreme cases, case A: $|Q| >> |U|$ and case B: $|Q| \sim |U|$.
For case A, the degree of polarization for single scattered photons of polarized incident photons is expressed  by using equation (\ref{deg-polz}) as  
\beqn\label{deg-polz-polA}
P_A = \frac{Q}{I} = \frac{2 - 2\sin^2\phi(1+\cos^2\theta)}{\frac{k'}{k}+\frac{k}{k'}-2\sin^2\phi\sin^2\theta}
\eeqn

For case B, it is expressed  by using equation (\ref{p-chi}) as  \cite[see, e.g.,][]{Matt-etal1996, Lei-etal1997}
\beqn\label{deg-polz-polB}
P_B =  \frac{\sqrt{Q^2+U^2}}{I} = \frac{2 - 2\sin^2\phi\sin^2\theta}{\frac{k'}{k}+\frac{k}{k'}-2\sin^2\phi\sin^2\theta}
\eeqn
 In Thomson regime, $P_B$ = 1.\\ 
Some interesting facts:
{\tt  \textbf{ (i) $\theta$ = 0:}} For $\theta = 0$, $Q \propto \cos(2\phi)$ and $U \propto \sin(2\phi)$.  
Since for $\theta=0$ one has $k=k'$ which gives $P_B$ = 1 for all $\phi$.  
 However, $U$ = 0 for $\phi = 0, \pi/2, \pi$, and in this case PD would be determined by $P_A$.  
Here, $P_A$ = 1 for $\phi = 0, \pi$, and $P_A$ = -1 for $\phi = \pi/2$.
Conclusively, $|P|$ = 1 for $\theta = 0$.
   {\tt \textbf{ (ii) $\phi$ = 0, $\pi/2$:}} For $\phi$ = 0 or $\pi/2$, 
   $U = 0$. The PD is  $P_A = \frac{2}{\frac{k}{k'}+\frac{k'}{k}}$ and  $\frac{-2\cos^2\theta}{\frac{k}{k'}+\frac{k'}{k}-2\sin^2\theta}$ for $\phi$ = 0 and $\pi/2$ respectively \cite[for $\phi =0$, see][]{McMaster-1961}.
   In Thomson limit,  $P_A $ = 1  and -1; $\chi$ = 0 and $\pi/2$ for $\phi = 0$ and $\pi/2$ respectively  (see also Figure \ref{fig:PD-pol}), also by definition of $Q$, here $\theta'_e$ = 0 and $\pi/2$ respectively. And in words, the polarization properties of scattered photons are same to the incident photons.
   {\tt \textbf{(iii) $\phi$ = $\pi$/4:}} For $\phi = \pi/4$, it is expected that the incident polarized photons behave like unpolarized photons. It means that the degree of polarization of scattered photons will be described by expression (\ref{deg-polz-un}). We note that for $\phi = \pi/4$ the only $P_A$ reduces to the expression (\ref{deg-polz-un}).
   
{\tt Modulation curve:} 
The expression for modulation curve is written by using equation (\ref{gen-dif-polz}) as
\begin{footnotesize}
\begin{align} \label{eq:modu_pol}
  \frac{d\sigma}{d\Omega}^{pol} = \frac{1}{4}r_o^2\left(\frac{k'}{k}\right)^2&\left[\frac{k}{k'}+\frac{k'}{k}-2 + 4(\cos^2\phi \cos^2\theta'_e + \sin^2\phi \sin^2\theta'_e \cos^2\theta \right. \nonumber \\
    & \qquad \left. - \frac{1}{2}\sin2\phi\sin2\theta'_e \cos\theta) \right].
\end{align} \end{footnotesize}
After rearranging the term, the above equation can be written as
\begin{footnotesize} \begin{align*}\frac{d\sigma}{d\Omega}^{pol} = \frac{1}{4}r_o^2\left(\frac{k'}{k}\right)^2& \left[\frac{k}{k'}+\frac{k'}{k}-2\sin^2\phi \sin^2\theta \right] \left[1 + P_A \left(\cos2\theta'_e - \right. \right. \nonumber \\
      & \left. \left. \frac{\sin 2\phi \cos\theta}{\cos^2\phi-2\sin^2\phi\cos^2\theta} \sin 2\theta'_e \right)\right]\end{align*}\end{footnotesize}
%%\beqn \label{eq:modu_pol}

\subsection{A special case for polarization measurement at $\theta$ = 0}
\label{theta_0}

For $\theta$ = 0, the cross section (\ref{gen-dif-polz}) can be written simply as
\beqn \label{scat0-dif-polz}
\frac{d\sigma}{d\Omega} = r_o^2 \cos^2(\theta_e+\theta'_e),
\eeqn
here we use  $\frac{k'}{k} = 1$, for $\theta$ = 0, which  is true for a electron is in either rest or motion.
In this case, $k'_e$ will also lie on the perpendicular plane to the $k$, 
and thus for a plane %angle-measurement
the solid angle becomes $d\Omega$ = $d\theta'_e$.
Hence, $d\sigma$ $\propto$ $\cos^2(\theta_e+\theta'_e) d\theta'_e$,
which has  %predicts
properties that the distribution of $\theta'_e$ replicates the polarized distribution of $\theta_e$. It can be understood as i) for unpolarized incident photons, $\theta_e$ is isotropically distributed thus the averaged cross section over $\theta_e$ for $\theta'_e$ is simply a constant, or
\[\frac{d\sigma}{d\theta'_e} = constant. \]
ii) for polarized incident photons, $\theta_e$ = constant = $\phi$, and the cross section becomes
\[\frac{d\sigma}{d\theta'_e} = r_o^2 \cos^2(\theta'_e+\phi) = \frac{r_o^2}{2} \left(1 + \cos(2(\theta'_e+\phi))\right),\]
which is a modulation curve for the scattered photons with degree of polarization $P$ = 1 and the angle of polarization $\chi = \phi$.

In general for the partially polarized incident photons, in which $P$ fraction is the polarized photons with polarization angle $\phi$ and $(1-P)$ fraction is the unpolarized photons, the modulation curve can be written as
\beqn \label{mod-polz-scat0}
\frac{d\sigma}{d\theta'_e} = A + B \cos(2(\theta'_e+\phi))
\eeqn
here, A and B are a normalization factor, and clearly, $P = B/A$ and $\chi = \phi$. The above expression is similar to the equation (4.10) of \cite{Lei-etal1997}.

\subsection{Lorentz invariance of the Stokes parameters}\label{l:inv}
We know that the field of the radiation is transverse in any reference frame, and the
 Lorentz boost subjects to an aberration effect of radiation. Since the electric vector always lies on the perpendicular plane to radiation propagation direction, and these electric vectors will be transformed from one frame to another Lorentz-boosted frame with the same rule.
Thus if the radiation is completely polarized in one reference frame, then 
it will be completely polarized in any Lorentz boosted  frame. 
In other words, the degree of polarization  of  photons in any
Lorentz-boosted frame is same to the magnitude of PD in the electron rest frame.
Later, we will argue that in Compton scattering the angle of polarization for
photons remains the same in any Lorentz-boosted frame.
Hence, the Stokes parameters are invariant under Lorentz transformations \cite[see, e.g.,][references therein]{Landau-Lifshitz1987, Krawczynski2012}. 

\section{Monte Carlo Method}\label{mc:method}
The Klein-Nishina differential cross section for unpolarized rest electrons expressed by equation (\ref{gen-dif-polz}) depends mainly on momentum of incident photon $k$, $\Theta$-angle and scattering angle $\theta$. For a given incident photon direction and polarization
angle $\phi$ (=$\theta_e$ by assumption), in principle, without affecting the cross section one can take
any direction of ($k \times k'$)  on the perpendicular plane to $k$ with maintaining the $k_e$ direction such that the angle between $k_e$ and ($k \times k'$)
is $\theta_e$. Hence, for a known incident photon direction and polarization, any scattering plane is permissible according to the cross section unless the direction of $k_e$ is not fixed in space (say, global coordinate).
In case of the fixed $k_e$ in space, the photon can scatter onto two planes only, as there are only two possible ways for the  ($k \times k'$) presentations, left and right side of the $k_e$.
Next, for a known incident photon direction,  polarization angle and fixed $k_e$ % polarization angle %$\theta_e$
in space (i.e., fixed scattering plane) if the $\Theta$-angle is known then according to the cross section the electric vector of scattered photon $k'_e$ lies on the surface of cone with opening angle $\Theta$ and cone-axis along the  $k_e$.
Since, the scattering plane is fixed, so $(k \times k')$ also. And the $k'$ direction can be determined in a perpendicular direction to the plane containing $k'_e$ and $(k \times k')$, in which $k'_e$ is any one of vectors which lies on that cone.
Simply, if one takes $(k \times k')$ as a $z$-axis then the intersection %direction
of $\phi$-plane and that cone gives $k'_e$ and the normal to this $\phi$-plane (which cuts that cone) will give $k'$ (see Figure \ref{fig:1ang}, however for clarity, the cone containing possible $k'_e$ is not shown).
It can be understood easily when $k_e$ lies on the scattering plane (i.e., ($x,y$)-plane), and here one can note that for a particular value of $\Theta$ either some definite range of $\theta$ or, $\theta'_e$ is possible. In another example when $k_e$ is along the ($k \times k'$), in this case, the photon can scatter in all possible directions of the scattering plane, and obviously $\theta'_e = \Theta$.
Hence for a given polarization characteristics of incident photon, and for given  $\Theta$-angle, only the definite range of  $\theta$ and  $\theta'_e$ is permissible, where $\theta$ and $\theta'_e$ both are related each other by  equation (\ref{cos:pol}).

There are mainly two unknown quantities $\Theta$-angle and scattering angle $\theta$ for determining the Klein-Nishina cross section, as we know $k$ and $k_e$ prior to scattering (at least, in MC calculation). But, to describe the scattered photon polarization properties one needs the angle $\theta'_e$, which can be obtained by using equation (\ref{cos:pol}) for a known $\Theta$-angle and $\theta$.
Therefore, to examine the polarization in Compton scattering, 
one have three unknown quantities $\Theta$-angle, $\theta$ and $\theta'_e$, in which two quantities would be extracted from Klein-Nishina cross section and
remaining one would be obtained by using equation (\ref{cos:pol}).
In above paragraph we note that in ($k \times k'$)-coordinate one can easily know the possible range of $\theta$ (or $\theta'_e$) value for a given $\Theta$-angle and $k_e$.
So, it is more convincing, if one  describe the Compton scattering  in ($k \times k'$) coordinate system (where the $z-$axis is along the ($k \times k') $ )
with expressing the cross section as a function of $\theta$ and $\theta'_e$ (see Figure \ref{fig:1ang}).
In the present study for Monte Carlo (MC) calculations we consider the cross section, equation (\ref{gen-dif-polz}), as a function of $\theta$ and $\theta'_e$ and describe the Compton scattering locally in ($k \times k'$)-coordinate system.
The algorithm for MC method is similar to the algorithm of \citep{Kumar-Misra2016b} with additional inclusion of %information about
polarization properties. Below we have described the important steps involved in MC calculations.

To describe the different steps involved in MC calculations we consider, for simplicity, a spherical corona of radius $L$ and temperature $T_e$. The seed photon source is situated at the origin of spherical corona which illuminates in all directions. The optical depth $\tau$ is defined along the radial direction of the corona/medium, the electron density inside the corona is $n_e = \frac{\tau}{L\sigma_T}$, where $\sigma_T$ is Thomson cross section.  
We assume that the seed photon source is a black body with temperature $T_b$.
We track a photon till it leaves the medium after single/multiple or zero scattering. We repeat the process for a large number of 
photons to make the statistics analysis. 
\begin{itemize}
\item In first step, we determine the incident photon's energy $E = h\nu$ from a black body distribution and the electron's velocity from the velocity distribution of temperature $T_e$. We consider an isotropic distribution for the photons and electrons direction. The mean free path $\lambda$ of photon of energy $E$ is computed for a given $T_e$ and $n_e$. We compute the above quantities in global coordinate using the  scheme developed by \cite{Hua-Titarchuk1995} (see also  \cite{Krawczynski2012} for the polarization scheme).
  
\item Next, we determine the collision free path of the photon $l_f$ in the medium with an exponential pdf (probability distribution function), $\exp\left(\frac{-l_f}{\lambda}\right)$, and obtain the condition for occurrence of scattering.
  If $l_f > L$ then the photon escapes the medium without scattering and for $l_f < L$ the scattering will be happened at distance $l_f$ from the origin (or in general, from the previous site of scattering) in direction of the incident photon.

\item Next, we specify the polarization properties of the incident photon locally in $(k\times k')$-coordinate of global (say, ($k\times k')_{\text{global}}$-coordinate). 
  For this we first assign the $(k\times k')$ direction on the perpendicular plane to %the incident photon direction
  $k$. For an unpolarized incident photon, we select the $(k\times k')$ direction uniformly, also %its polarization angle
  $\theta_e$ angle uniformly to determine the polarization vector. For a polarized photon, we first fix  $k_e$ %the polarization vector
  and then determine the $(k\times k')$ direction either left or right side of $k_e$ at angle $\theta_e$ on the perpendicular plane to $k$. 

\item In second step to describe the Compton scattering, we transform the quantities from global coordinate to electron rest frame. %with $(k\times k')$-coordinate.
  \begin{itemize}
\item As the Stokes parameters are invariant under the Lorentz transformation, we assume that the polarization angle also does not change. In electron rest frame, due
  to the aberration effect the incident photon direction  will change in the plane containing $k\ \& \ \text v$ (say, $k_{ab}$). Consequently, the polarization vector will lie now in a plane perpendicular to $k_{ab}$, denoted as $k_{e}^{ab}$. The direction of $k_{e}^{ab}$ can be determined as. Since the scattering plane is assigned in global coordinate, so to determine the scattering plane in electron rest frame, we consider an another incident photon direction on global scattering plane and transformed it into electron rest frame, thus the plane containing $k_{ab}$ and this transformed photon direction would be the scattering plane in electron rest frame.
  Therefore, ($k\times k'$) in electron rest frame can be determined, and consequently one can fix the $k_{e}^{ab}$ for a known $\theta_e$ in this ($k\times k')_{\text {rest frame}}$-coordinate. With having $\theta'_e$ and $\theta$ from the cross section in electron rest frame one can determine the scattered photon direction $k'_{ab}$ and its $k_e^{'ab}$. In a similar way, these two quantities transferred back to the ($k\times k')_{\text {global}}$-coordinate. We again emphasize that one have to transform the $k_e^{'ab}$ into $k_e^{'}$ with the condition of $\left. \theta'_e \right|_{\text{lab frame}} = \left. \theta'_e \right|_{\text{rest frame}}$ due to the  Lorentz invariance of Stokes parameters.

\item The condition of $\left. \theta_e \right|_{\text{lab frame}}= \left. \theta_e \right|_{\text{rest frame}}$ can be understood as. Suppose, the incident photons are fully polarized with $\theta_e$ = 45$\degree$. The degree of polarization for scattered photons with almost rest electron will be described by the equation (\ref{deg-polz-polA}) for $\phi$ =45$\degree$ (see also Figure \ref{fig:PD-pol}). 
  Since the degree of polarization is a Lorentz  invariant quantity. 
  Therefore, if these completely polarized photons scatter with moving electron then PD will still describe by equation (\ref{deg-polz-polA}) for $\phi$ =45$\degree$, which can be only possible when $\left.\theta_e\right|_{lab frame} = \left. \theta_e \right|_{\text{rest frame}}$. Hence, %As stated earlier,
  in Compton scattering process the PD and $\theta_e$ both are invariant over the Lorentz-boosted frame.

\end{itemize}
\item As the cross section depends only $\theta'_e$ 
we skip the all steps which involve to determine the $k_e^{ab}$, $k_e^{'ab}$ and $(k \times k')_{\text{rest frame}}$,  
and simply extract  $\theta$ and  $\theta'_e$ from the cross section in electron rest frame.
\item Next, we compute the scattered photon frequency (using equation \ref{eq:sctfq}), and transform back this frequency to global (lab) coordinate by computing the angle between scattered photon and incident electron. 
In addition, after the reverse aberration effect the scattered photon has to lie on the pre-defined scattering plane. We first compute the scattered photon's propagation direction ($k'$) and polarization vector $k'_e$ (i.e., on perpendicular plane to $k'$) in local $(k\times k')_{\text{global}}$-coordinate and then transform back to those in the global coordinate.

\item Next, we  estimate the collision free distance $l_f$ for the scattered photon (of energy $E' = h\nu'$) and find the distance of next site of scattering from the origin, say $r_n$. If $r_n < L$ then next scattering will occur otherwise photon will escape the medium.
\item For next scattering, we first determine the angle of polarization, say $\chi_s$ of scattered photon using equation (\ref{av_pol}). 
  Since for double scatterings or two consecutive scatterings the angle between previous and next scattering planes is $\chi_s$ \cite[e.g.,][]{McMaster-1961}, so using this we compute the $(k \times k')_{next}$ for next scattering on perpendicular plane to the $k'$.
  And,  for a known $k'_e$ in global coordinate we compute the polarization angle $\theta_e$ wrt $(k \times k')_{next}$. We proceed the calculations for next scattering with treating $k'$ of previous scattering as an incident photon, and follow the same steps until the scattered photon escapes the medium. 
\end{itemize}

\section{MC Results verification}\label{sec:verif}

We verify the polarization results of MC calculations 
with a theory 
which is revisited in section \S\ref{sec:2}, i.e., for single scattering, and almost rest electron. Since, the theoretical results are derived for a given scattering plane, so in MC calculation we obtain the results without bothering about orientation of the scattering plane.
However, later we consider the orientation of the scattering plane, see section \S\ref{sec:Chandra}. %, in computations of PD and PA.
In the following sections we show the MC results for polarized/ unpolarized incident photons. But we will first discuss the way of computing the PD and PA for scattered photons with arbitrary scattering numbers using the results of section \S\ref{theta_0}.

\subsection{General modulation curve to estimate the PD $ \& $ PA}\label{sec:gen-modu-p}

In section \S\ref{theta_0}, we showed that one can know the polarization properties of
incident photons $P \ \& \ \chi$ by mapping the distribution of  $\theta_e'$ of scattered photons at $\theta$ $\sim$ 0 after single scattering.
In general, one can estimate the polarization properties of scattered photons with any average scattering no., say $\langle N_{sc} \rangle$ by mapping the distribution of  $\theta_e'$ of scattered photons of $\langle N_{sc} + 1 \rangle$ scattering no. at $\theta$ $\sim$ 0. 
Mathematically, we are essentially using here a probability $\propto \cos^2(\theta_e|_{\langle N_{sc}+1\rangle} + \theta'_e|_{\langle N_{sc}+1\rangle})$ for constructing the distribution of $\theta'_e|_{\langle N_{sc}+1\rangle}$ for a known $\theta_e|_{\langle N_{sc}+1\rangle}$ (see equation \ref{scat0-dif-polz}). From equation (\ref{chi_th_next}), we can write $\theta_e|_{\langle N_{sc}+1 \rangle} = (\theta'_e \pm \chi)|_{\langle N_{sc}\rangle}$, here the subscript with vertical bar is used for the quantity related to that scattering number, $\chi$ is computed using equation (\ref{av_pol}).
Hence, now without going for the calculations of $\langle N_{sc} + 1\rangle$th scattering  we can estimate the polarization properties of $\langle N_{sc}\rangle$th scattered photons by constructing the modulation curve of $\eta$-angle for a known value of $(\theta'_e \pm \chi)|_{\langle N_{sc}\rangle}$ using the probability $p(\eta)$ 
\beqn \label{gen-modu-p} 
p(\eta) \propto \cos^2(\eta+(\theta'_e \pm \chi)|_{\langle N_{sc}\rangle}) d\eta.
\eeqn
We have used this $p(\eta)$ in the MC calculations to estimate the polarization properties of the scattered photons.

\begin{figure}
\centering
\begin{tabular}{c}\hspace{-1.5cm}
  \includegraphics[width=0.37\textwidth]{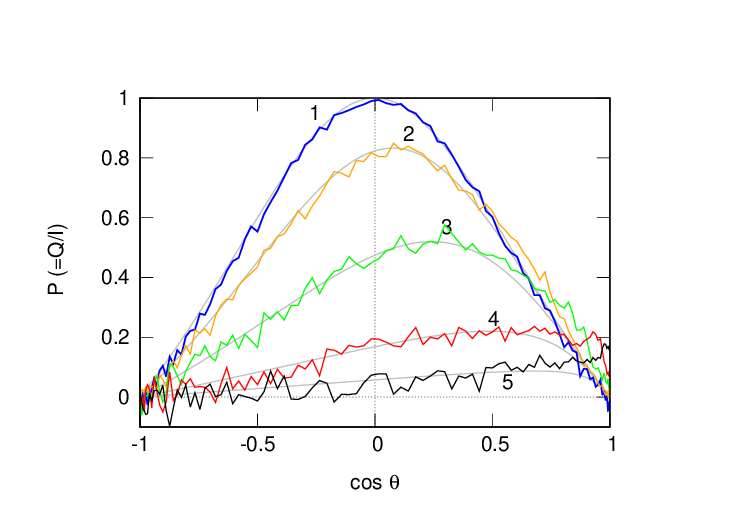}  \\%m-unp-dp-thn.eps
\end{tabular}\vspace{-0.3cm}
\caption{The degree of polarization after single scattering of unpolarized (incident) photons as a function of scattering angle for five different photon energies. Here the curves 1, 2, 3, 4 and 5 are for incident photon energies 3, 300, 900, 3000 and 9000 keV respectively. The gray solid curves are analytic one expressed by equation (\ref{deg-polz-un}).  }
\label{fig:PD-unpolz}
\end{figure}

\begin{figure*}%[h!]\vspace{-0.19cm}
%%\captionsetup{font=footnotesize}
\centering
\begin{tabular}{lll}%\hspace{-1.5cm}
  \includegraphics[width=0.37\textwidth]{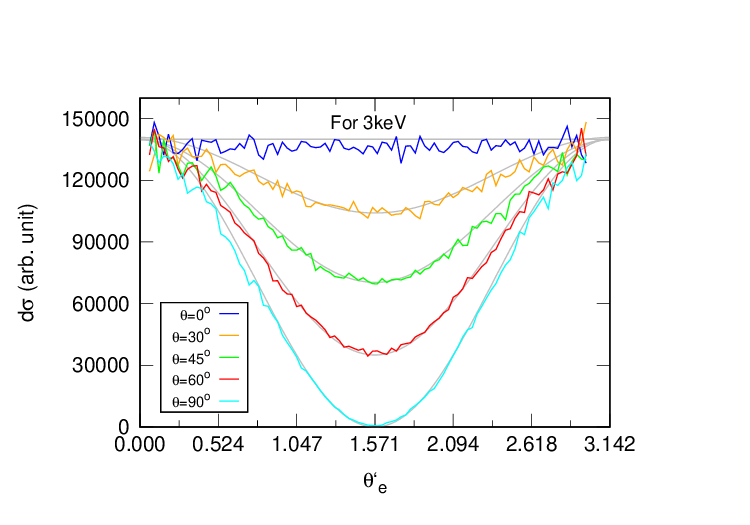}&\hspace{-1.5cm}%m3pl-sig-mod-thn.eps
  \includegraphics[width=0.37\textwidth]{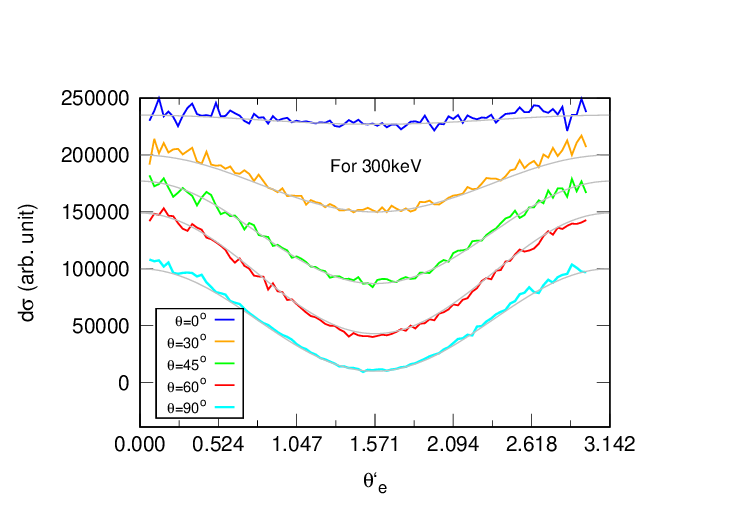}&\hspace{-1.5cm}%m300pl-sig-mod-thn.eps
   \includegraphics[width=0.37\textwidth]{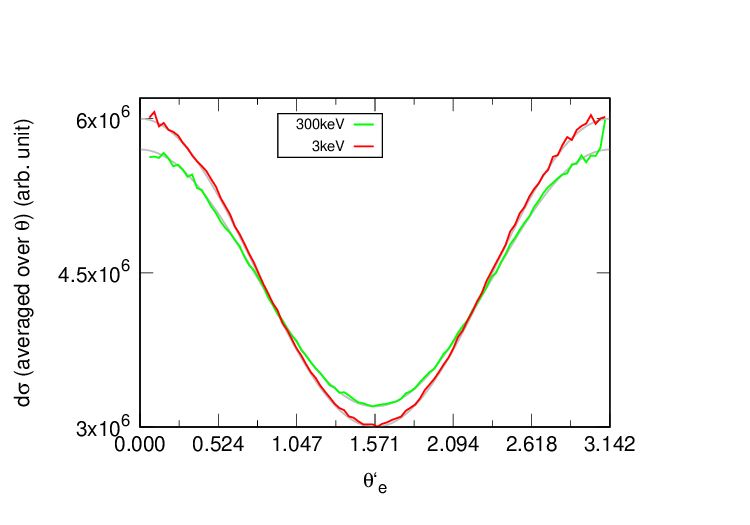}\\%m-upl-sig-avg-modn.eps
\end{tabular}\vspace{-0.3cm}
\caption{The modulation curves after single scattering of unpolarized (incident) monochromatic  photons. The left and middle panels are for the fixed $\theta$ and for the incident photon energy 3 and 300 keV respectively  while the right panel is for the averaged $\theta$. In left and middle panels the blue, orange, green, red and cyan curves are for $\theta$ = 0, 30, 45, 60 and 90 degree respectively. The gray solid curves are the analytical one expressed by equation (\ref{mod_un}) for a given $\theta$. In right panel, the red and green curves are for photon energies 3 and 300 keV respectively. The gray curves for 3 and 300 keV are $4.5\times10^6(1+0.33\cos 2\theta'_e)$ and $4.45\times 10^6(1+0.28\cos 2\theta'_e)$ respectively, which reflects the PD = 0.33 and 0.28 by comparing the equation (\ref{mod-polz-scat0}) for 3 and 300 keV respectively.   } 
\label{fig:sigm-unpol}
\end{figure*}

\subsection{MC results for unpolarized incident photons}
In Figure \ref{fig:PD-unpolz} we show the PD as a function of scattering angle for the unpolarized monochromatic incident photons for five different energies 3, 300, 900, 3000, 9000 keV. The solid gray curves are for analytic PD expressed by equation (\ref{deg-polz-un}),  the MC results are consistent with analytic ones. % are matched with each other.
Clearly in Thomson regime, at $\theta =$ 90$\degree$ the single scattered unpolarized (incident) photons are completely polarized.
In Figure \ref{fig:sigm-unpol}, the modulation curves for a given $\theta$ have been shown for two different photon energies 3 (in left panel) and 300 (in middle panel) keV. In both cases, the MC results match with analytic one, equation (\ref{mod_un}) for a given $\theta$ which is shown by gray curve.
By comparing with equation (\ref{mod-polz-scat0}) for all curves of left and middle panels we have $\phi$ = 0$\degree$, which signifies that the scattered photons are polarized in a perpendicular direction to the scattering plane.
In the right panel we show the averaged modulation curve over $\theta$ for two photon energies 3 and 300 keV. The gray curves for 3 and 300 keV are for equation $4.5\times 10^6(1+0.33\cos 2\theta'_e)$ and $4.45\times 10^6(1+0.28\cos 2\theta'_e)$ respectively. By comparing with equation (\ref{mod-polz-scat0}) the estimated PD of single scattered 3 and 300 keV unpolarized photons are $\sim$0.33 and 0.28. Since we know the distribution of  $\theta$ and know the PD as a function of $\theta$, we have computed the averaged PD weighted over the $\theta$. And we find the PD for 3 and 300 keV photons  are $\sim$0.28 and 0.24 respectively, thus the both methods almost agree with each other. In addition, for 3 keV photons (which is in Thomson regime) we compute the averaged PD analytically as $\langle P \rangle = \frac{\int_0^\pi P (1+ \cos^2\theta) d\theta}{\int_0^\pi (1+\cos^2\theta) d\theta}$ = 1/3 with considering $d\sigma \propto (1+\cos^2\theta)$ in Thomson regime.
Here we reemphasize that the high value of PD of single scattered unpolarized (incident) photons is due to the fixed scattering plane (see also case I with $\theta_i \equiv [0, \pi]$ of section \S\ref{sec:Chandra}) and if one accounts the effect of orientation of the scattering plane, then the  PD magnitude will get reduced, see the section \S\ref{sec:Chandra} for details.  

\begin{figure}%[h!]\vspace{-0.19cm}
%%\captionsetup{font=footnotesize}
\centering
\begin{tabular}{ll}\hspace{-1.cm}
 \includegraphics[width=0.32\textwidth]{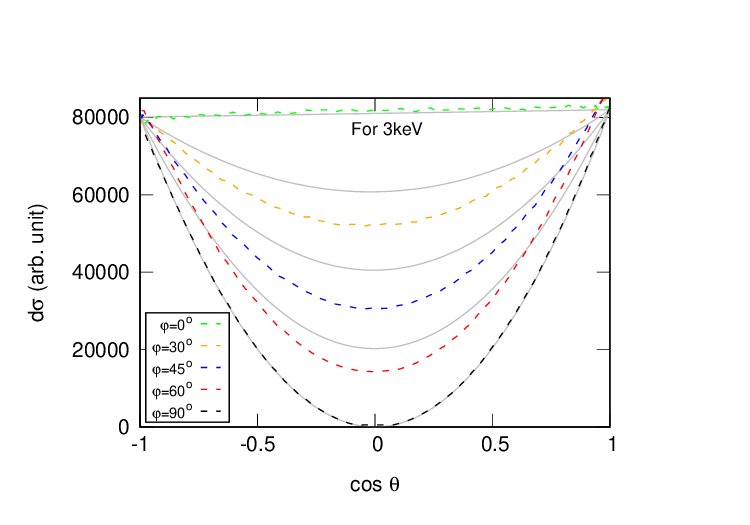}&\hspace{-1.5cm}% m3pl-phi-sig-thn.eps
  \includegraphics[width=0.32\textwidth]{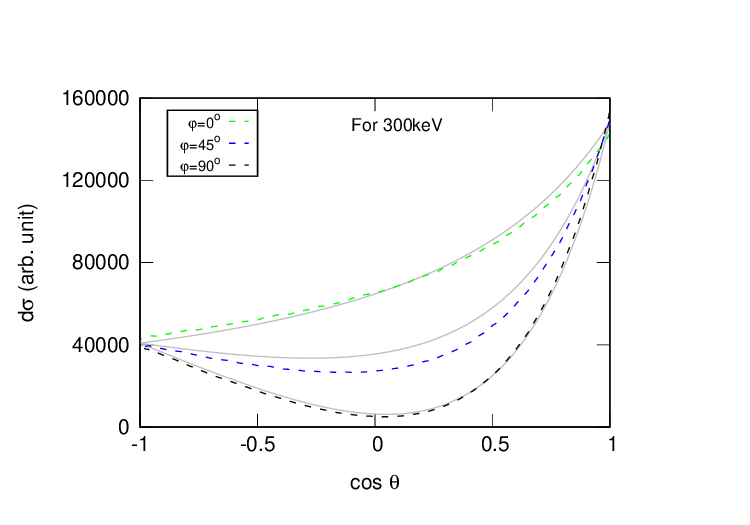}\\%m300pl-phi-sig-thn.eps  
\end{tabular}\vspace{-0.3cm}
\caption{The Klein-Nishina cross section as a function of $\theta$ for a completely polarized incident photon of polarization angle $\phi = \theta_e$. The dashed curves are calculated one by MC method using the equation (\ref{gen-dif-polz}) while the gray solid curves are analytic one expressed by equation (\ref{eq:cros_pol}) and the observed mismatched is due to the approximation involved in equation (\ref{eq:cros_pol}), see the text for details. The left and right panels are for incident photon energy 3 and 300 keV respectively,  and the dashed green, orange, blue, red and black curves are for $\theta_e$ = 0, 30, 45, 60 and 90 degree respectively. }
\label{fig:sigm-pol}
\end{figure}

\subsection{MC results for polarized incident photons}
In section \S\ref{sec2:pol}, we argue that the cross section for a completely polarized incident photon of polarization angle $\phi$ is obtained with an assumption of isotropic distribution of $\theta'_e$. We examine this assumption in MC calculations. We compute the averaged cross section over $\theta_e'$ as a function of $\theta$ (using the equation \ref{gen-dif-polz}) for photon energies 3 and 300 keV, the results are shown
in Figure \ref{fig:sigm-pol}. 
We find that the computed cross sections are significantly deviated from the equation (\ref{eq:cros_pol}) except for $\phi$ $\sim$0 and 90$\degree$, where it shows a slight deviation.
  Actually in case of $\phi$ $\sim$0 and 90$\degree$, the equation (\ref{gen-dif-polz}) nearly reduces to the equation (\ref{eq:cros_pol}), as here also $\theta'_e$ = 0 and 90$\degree$ (see the facts (ii) of section \S\ref{sec2:pol}) for $\phi$ $\sim$0 and 90$\degree$ respectively.
  Hence,
the distribution of $\theta_e'$-angle is no longer isotropic (e.g., \cite{Matt-etal1996}), as stated in the previous section.

\begin{figure}%[h!]\vspace{-0.19cm}
%%\captionsetup{font=footnotesize}
\centering
\begin{tabular}{c}%\hspace{-1.5cm}
  \includegraphics[width=0.37\textwidth]{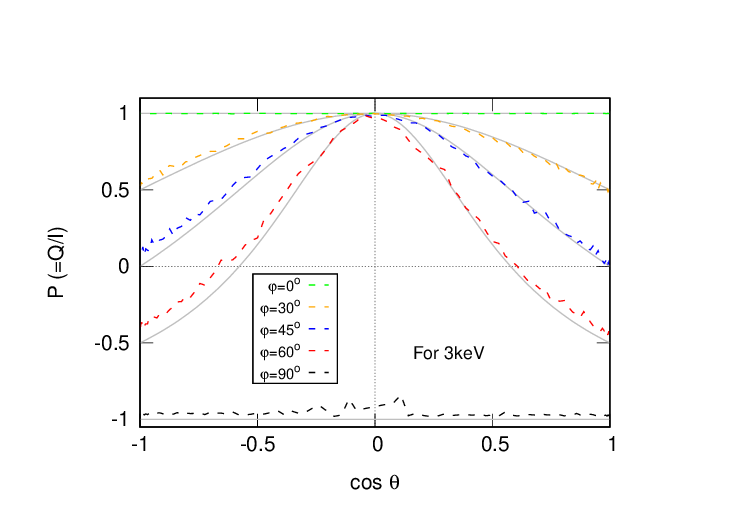}\\%m3pl-phi-dp-thn.eps 
\end{tabular}\vspace{-0.3cm}
\caption{The degree of polarization after single scattering of completely polarized (incident) photons of energy 3 keV as a function of scattering angle for five different polarization angle. Here, the dashed green, orange, blue, red and black curves are for $\theta_e$ = 0, 30, 45, 60 and 90 degree respectively, and the solid gray curves are an analytical one expressed by equation (\ref{deg-polz-polA}).   }
\label{fig:PD-pol}
\end{figure}

\begin{figure}%[h!]\vspace{-0.19cm}
%%\captionsetup{font=footnotesize}
\centering
\begin{tabular}{c}\hspace{-1.5cm}
  \includegraphics[width=0.4\textwidth]{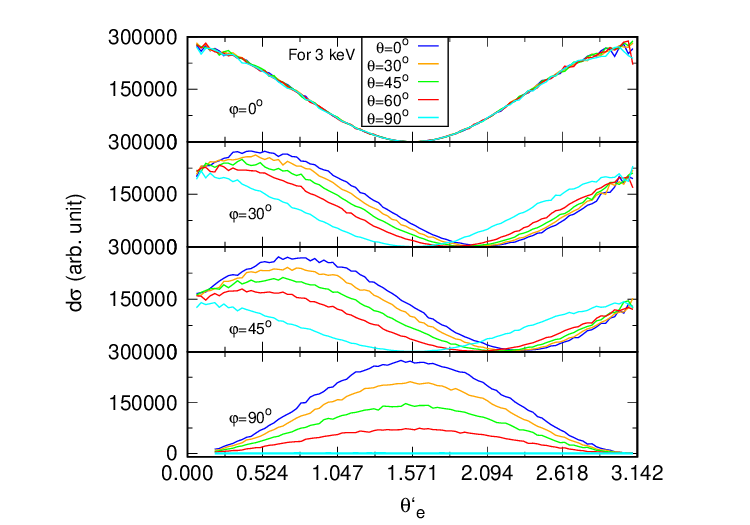} %m3pl-modu_sctn.eps 
\end{tabular}\vspace{-0.3cm}
\caption{The modulation curves after single scattering of completely polarized
  (incident)  photons of energy 3 keV for a given $\theta$. 
  The solid blue, orange, green, red and cyan curves are for $\theta$ = 0, 30, 45, 60 and 90 degree respectively. %In left and right panels,
  The first (top), 2nd, 3rd and 4th (last) panels are for $\theta_e$ = 0, 30, 45 and 90 degree respectively. The curves are described analytically by equation (\ref{eq:modu_pol}).  }
\label{fig:4mod-pol}
\end{figure}

In Figure \ref{fig:PD-pol}, we show the PD as a function of $\theta$ for single scattered completely polarized photons of energy 3 keV. We find that the MC results agree with the
corresponding theoretical PD (shown by gray solid curve) expressed 
by equation (\ref{deg-polz-polA}). Here we remind that the positive value of PD signifies that the scattered photons polarization is  perpendicular to the scattering plane while the 
polarization is along the scattering plane for negative value.   
In Figure \ref{fig:4mod-pol}, we show the modulation curves of single scattered polarized photons for a fixed $\theta$. We have computed the results for four different polarization angles of incident photons 
of energy 3 keV.
For $\phi$ = 0, the equation (\ref{eq:modu_pol}) predicts that the modulation curve is independent of $\theta$ in Thomson regime, we find same here 
see the top left panel. In general, these modulation curves will be described by the equation (\ref{eq:modu_pol}).


\section{Modulation curve in perpendicular plane of incident photon for $\theta <$ 25$\degree$: a detector working principle}\label{sec:modu}

In this section, we examine the result of section \S\ref{theta_0}, and are interested to find the range of $\theta$ in which the polarization properties of
incident photons can be derived from the modulation curve of scattered photons.
Motivation for this exercise is that the cross section is maximum for $\theta = 0$ for polarized/ unpolarized photons of any energy (see Figure \ref{fig:sigm-pol}, \ref{fig:sigm-unpol} respectively). Thus it is expected practically that
on average upto some range of $\theta$ ($> 0$), the polarization properties of scattered photons at $\theta = 0$ will dominant over that range of scattered photons. 
  We find that for $\theta < 25\degree$ (or $\theta \equiv [0,25\degree]$) one can adequately predict the polarization properties of $k$ with using the  modulation curve for the $k'$. 
  
\begin{figure}%[h!]\vspace{-0.19cm}
%%\captionsetup{font=footnotesize}
\centering
\begin{tabular}{l}
  \includegraphics[width=0.27\textwidth]{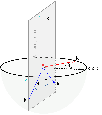}%fig7f.eps
\end{tabular}\vspace{-0.3cm}
\caption{A schematic diagram for calculating the modulation curve in perpendicular plane of incident photon. Here, $k$ is along the $z$-axis, $(k \times k')$ is along the $x$-axis, thus the ($z,y$)-plane is a scattering plane. $k_e$ with fixed $\theta_e$ and $k'$ for scattering angle $\theta$ are shown. }
\label{fig:2detecto}
\end{figure}

\begin{figure}%[h!]\vspace{-0.19cm}
%%\captionsetup{font=footnotesize}
\centering
\begin{tabular}{l}
\includegraphics[width=0.38\textwidth]{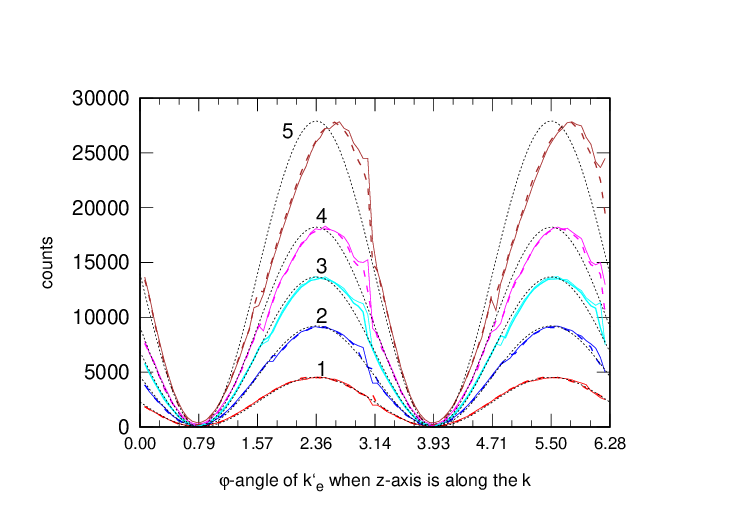}\vspace{-1.cm} \\% mo3-300-pl45-detc.eps
\includegraphics[width=0.38\textwidth]{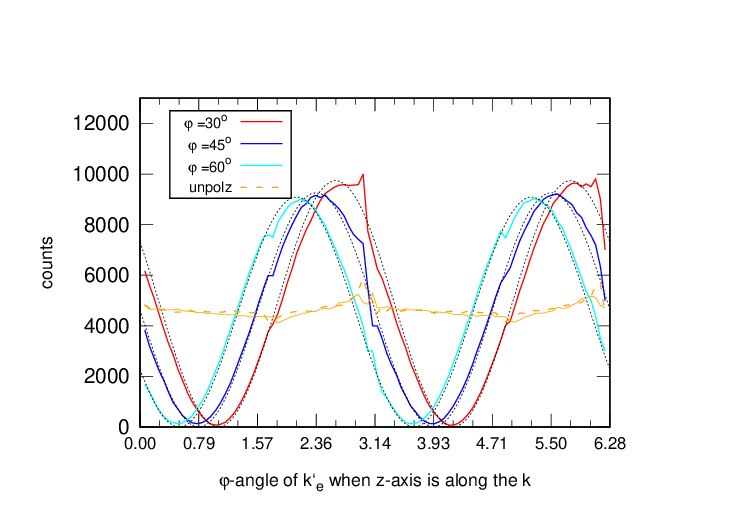} %mo3-pl-ang-unpn.eps 
  \end{tabular}\vspace{-0.45cm}
\caption{$\phi$-angle distribution of $k'_e$ after single scattering when the completely polarized photons incident along the $z$-axis. The upper panel is for the polarization angle $\theta_e$ = 45$\degree$, and the curves 1, 2, 3, 4 and 5 are for the different ranges of $\theta$ $\equiv$ [0,7.5$\degree$], [0,15$\degree$], [0,22.5$\degree$], [0,30$\degree$] and [0,45$\degree$] respectively. 
  The lower panel is for three different $\theta_e$ at a given range of $\theta$ $\equiv$ [0,15$\degree$], where the solid red, blue and cyan curves are for $\theta_e$ = 30, 45 and 60$\degree$ respectively and the orange curves are for unpolarized incident photons. In both panels, the solid and dashed curves are for incident photon energy 3 and 300 keV respectively and the dotted curve is simply $a(\cos(2(\phi-\theta_e))+1)$, a is the normalization factor. }
\label{fig:work-detec}
\end{figure}

In view of practice, we also examine the same by constructing the distribution of projection of $k'_e$ on the perpendicular plane to $k$.
For this we
fix the direction of the incident photon, say $k$ lies along the $z$-axis. For a polarized photon, the direction of the electric vector is fixed in space, 
so ($k \times k'$) is also fixed. Without loss of generality, we consider ($k \times k'$) along the $x$-axis, that is the scattering plane is a $(z,y)$-plane. It is shown in Figure \ref{fig:2detecto}.
Clearly for $\theta = 0$, the $k'_e$ will lie on the $(x,y)$-plane. 
For $\theta > 0$, the $k'_e$ will not always lie on $(x,y)$-plane.
%For convenient,
And, we measure the direction of projection $k'_e$ on ($x,y)$-plane or the $\phi$-angle of $k'_e$. The distributions of $\phi$-angle of $k'_e$ for 3 and 300 keV completely polarized incident
photons have been
shown in upper panel of Figure \ref{fig:work-detec} for polarization angle $\phi$  = 45$\degree$ for five different ranges of $\theta$ = [0,7.5$\degree$], [0,15$\degree$], [0,22.5$\degree$], [0,30$\degree$],  [0.45$\degree$]. We also map the distribution of the $\phi$-angle of
$k'_e$ for three different polarization angles $\phi$  =  30$\degree$, 45$\degree$ and 60$\degree$ for a fixed $\theta$ range [0,15$\degree$], which is shown in the lower panel of Figure \ref{fig:work-detec}.
We find that for $\theta$ range $\approx [0,25\degree]$  %modulation curve
the distribution of projection of $k'_e$ on the  perpendicular
plane to $k$ after single scattering can estimate the polarization properties of incident photons $k$ adequately.

As understood, the detector has to estimate the polarization properties of the
observed photons. And, for Compton/ Thomson scattering based detector (like, AstroSat/CZTI, PoGo+, POLAR, POLIX etc \cite{Fabiani-2018}) the observed photons are  essentially an incident photon, and so the above analysis is more relevant for a detector as a working principle.
However, the above discussed are only the essential criteria  for the detector designing
and for the general mechanism for a specific detector please see the relevant references, like \cite[][and references therein]{Lei-etal1997, Fabiani-2018, Chattopadhyay-etal2014}. 

\section{Polarization measurement along a meridian plane after single scattering}\label{sec:Chandra}

We aim to estimate the degree of polarization and angle of polarization for the emergent photons from a given meridian plane after single scattering where unpolarized  incident photons lie on any appropriate plane. 
For this we consider a simplistic (without loss of generality) geometry a semi-spherical shell where the incident photon generates at center of the shell and we specify its direction in spherical coordinate by ($\theta_i, \phi_i$), i.e., on meridian $\phi_i$-plane. We are interested to compute the $P$ and $\chi$ for the scattered photon ($\theta_s, \phi_s$), or in general, to compute the $P\ \&\ \chi$ as a function of $\theta_s$ on the meridian $\phi_s$-plane.

Before discussing the general results, we first discuss, for clarity, a few simplistic cases in terms of incident photon direction ($k$), or scattering plane (or $(k \times k')$), or combination of both in the following sections.
To measure the angle of polarization one needs a reference direction of ($k \times k'$) for the next scattering, % (in detector),
we consider that ($k \times k'$)-next %direction which
lies on the same scattered meridian plane, and we take its direction as ($\theta_s+\pi/2,\phi_s$). 
However, we will first discuss the way of computing the averaged PD and PA of the scattered photons, as we are caring now the orientation of the scattering plane.

\subsection{Averaged degree of polarization}\label{avg_pd-pa}
The partially polarized incident photons of degree of polarization $p_I$ and the angle of polarization $\phi_I$ can be an average behaviour of total n type of incident photons which have different degree and angle of polarization, say $p_j$ and $\phi_j$ for $j^{th}$ type incident photons respectively. In context of Compton scattering, it can be expressed for $\xi$-angle variable as (using equation \ref{mod-polz-scat0})
\beqn \label{avg_dpa}
p_I \cos(2(\xi+\phi_I)) + 1 = \sum_{J=1}^n \frac{1}{n}[p_j \cos(2(\xi+\phi_j)) + 1 ],
\eeqn
For example, one has unpolarized photons which is a combination of two types of photon with  $p_1 =p_0, \phi_1 =\phi_0$ and $p_2 = p_0$, $\phi_2=\phi_o+\pi/2$. It would be true for any value of $p_0$ (e.g., $p_0 = 1$).
We use this characteristic to estimate the PD and PA for emergent photons either after single scattering or multi scattering. For this we compute the $\chi_j$ (PA of the $j^{th}$ type scattered photons) using equation (\ref{av_pol}), as $\tan 2\chi_j = \frac{U}{Q}$, and compute the averaged $p_j$ (over $\phi$, PA of the incident photons)  as a function of $\theta$ as, $p_j$ = $\frac{\langle Q\rangle}{\langle I\rangle}$.

\subsection{Case I: Fixed $k$ direction (along z-axis) for a given scattering plane}
We first consider a simplistic situation in which the unpolarized photons incident along the $z$-direction and we fix the scattering plane to $\phi_s$-plane, i.e., the ($k\times k'$) direction is ($\pi/2, \phi_s+\pi/2$). Since, the scattering plane is
the same as the interested meridian  $\phi_s$-plane, thus the scattering angle is $\theta = \theta_s$. And so $P$ will follow  equation (\ref{deg-polz-un}) with $\theta = \theta_s$. The $\chi$ is always 90 degree (with the choice of  ($k\times k'$)-next), it means that the $k'_e$ lies 
parallel to the $(x,y)$-plane. The result (for $\theta_i$=0) is shown in the 2nd column of table \ref{tab-I-II}.
In addition, similarly, for a fully polarized incident photons the PD will be determined by equation (\ref{deg-polz-polA}).

Instead of taking incident photons along the $z-$axis (or $\theta_i=0$), we also consider any direction of $k$ on incident meridian $\phi_i-$plane, i.e., the range of $\theta_i$ is $\theta_i \equiv [0,\pi]$. And we assume again that the scattering plane is the same as the interested meridian  $\phi_s$-plane. 
In this case, we expect a constant $P$ $\sim$0.33, as shown by the red curve in the right panel of Figure \ref{fig:sigm-unpol}. % over  $\theta_s$ variation.
We find almost the same and show the results in the 3rd column of table \ref{tab-I-II}.

\begin{table*}\vspace{-1.50cm}%[h!]
%%\captionsetup{font=footnotesize}
\centering
\caption{The PD ($P$) and PA ($\chi$) of the emergent photons as a function of $\theta_s$ (or $i$) for cases I and II. The results for case II are presented for six different scattered meridian planes $\phi_i + \phi_{is}$ with $\phi_{is}$ = 15, 30, 45, 60, 75 and 90 degree. While the results for case I are true for any scattered meridian plane, as here $\phi_s$-plane = $\phi_i$-plane, and computed for two different ranges of $\theta_i$ = 0 and [0,$\pi$].  }
\begin{footnotesize}
%\begin{tabular}{|p{0.80cm}p{1.3cm}p{1.3cm}*{6}{p{1.52cm}}| }
\begin{tabular}{|p{0.80cm}p{1.3cm}p{1.3cm}p{1.52cm}p{1.52cm}p{1.52cm}p{1.52cm}p{1.52cm}p{1.52cm}| }
  \hline
   & \multicolumn{8}{|c|}{$P \ \ (\chi)$} \\ \cline{2-9}
 $\cos(\theta_s)$ & \multicolumn{2}{|c}{\hspace{-0.2cm}Case I} & \multicolumn{6}{|c|}{Case II  \ $\phi_{si} ^1$ = (in degree)} \\ \cline{4-9}
  & \multicolumn{1}{|l}{$\theta_i$=0}& \multicolumn{1}{c|}{$\theta_i \equiv [0,\pi]$} & \multicolumn{1}{|l}{15$\pm$2} & 30$\pm$2 & 45$\pm$2 & 60$\pm$2 & 75$\pm$2 & 90$\pm$2 \\ \hline
  0.0 & 0.99 (90.9) & 0.28 (90.9) & 0.036 (180.0) & 0.137 (180.0) & 0.315 (178.1) & 0.569 (178.1) & 0.844 (176.3) & 0.660 (134.5) \\
  0.1 & 0.96 (90.9) & 0.27 (90.9) & 0.030 (178.1) & 0.132 (176.3) & 0.314 (174.5) & 0.570 (170.9) & 0.833 (160.0) & 0.948 (100.0) \\
  0.2 & 0.90 (90.9) & 0.27 (90.9) & 0.029 (176.3) & 0.128 (172.7) & 0.302 (169.0) & 0.543 (160.0) & 0.787 (143.6) & 0.905 (94.5) \\
  0.3 & 0.81 (90.9) & 0.26 (90.9) & 0.027 (174.5) & 0.120 (170.9) & 0.280 (163.6) & 0.499 (152.7) & 0.715 (130.9) & 0.815 (92.7) \\
  0.4 & 0.70 (90.9) & 0.26 (90.9) & 0.025 (174.5) & 0.109 (167.2) & 0.253 (158.1) & 0.442 (145.4) & 0.623 (123.6) & 0.704 (92.7) \\
  0.5 & 0.58 (90.9) & 0.26 (90.9) & 0.022 (172.7) & 0.096 (163.6) & 0.219 (152.7) & 0.375 (140.0) & 0.519 (118.1) & 0.581 (92.7) \\
  0.6 & 0.45 (90.9) & 0.26 (90.9) & 0.019 (170.9) & 0.080 (160.0) & 0.180 (149.0) & 0.302 (134.5) & 0.409 (114.5) & 0.455 (90.9) \\
  0.7 & 0.32 (90.9) & 0.26 (90.9) & 0.015 (169.0) & 0.062 (158.1) & 0.137 (145.4) & 0.224 (129.0) & 0.298 (110.9) & 0.328 (90.9) \\
  0.8 & 0.20 (90.9) & 0.26 (90.9) & 0.010 (167.2) & 0.042 (154.5) & 0.091 (141.8) & 0.147 (125.4) & 0.191 (109.0) & 0.210 (90.9) \\
  0.9 & 0.09 (90.9) & 0.27 (90.9) & 0.005 (165.4) & 0.021 (152.7) & 0.045 (138.1) & 0.070 (121.8) & 0.090 (107.2) & 0.098 (90.9) \\
  1.0 & 0.00 (90.9) & 0.29 (90.9)  & 0.000 (165.4) & 0.000 (150.9) & 0.000 (134.5) & 0.000 (120.0) & 0.000 (105.4) & 0.000 (90.9) \\
  \hline
\multicolumn{9}{l}{1: here $\phi_{si}$ is measured in anti-clockwise direction, if it is measured in clockwise direction then the angle of polarization becomes $\pi-\chi$} \\  
\end{tabular}\vspace{0.3cm}\end{footnotesize}
\label{tab-I-II}
\end{table*}

\subsection{Case II: ($k \times k'$) lies on the meridian plane of $k$}
In the previous case, the scattering plane and interested meridian plane are the same.
Next we consider that incident photons 
and
the ($k \times k'$) both  lie on the same  $\phi_i$-meridian plane, so ($k \times k'$)'s direction for a given $k$ 
is ($\theta_i+\pi/2, \phi_i$). Now the photons will scattered any meridian ($\phi_i+\phi_{is}$)-plane, in which $\phi_{is}$ is the angle between incident and scattered meridian plane and the range of $\phi_{is}$ is $\equiv$ [$-\pi, \pi$]. 
For $\theta_i$ = 0 and $\pi/2$, the scattering plane of $k$ is $(\phi_i \pm \pi/2)$-plane  and $(\theta = \pi/2)$-plane respectively. In general,
the scattering plane of $k$ will must pass the line 
($\pi/2,\phi_i+\pi/2$) (i.e., rotated $y$-axis with angle $\phi_i$ about $z$-axis, say $y'$-axis), or in other words it is a rotation of $(\phi_i \pm \pi/2)$-plane about $y'$-axis with same angle $\theta_i$.
Thus, except for scattered meridian $(\phi_i \pm\pi/2)$-plane, 
the $k'$ arises uniquely on a different scattering plane with different $\theta$.
For example, for the meridian $(\phi_i+\pi/6)$-plane  the $\theta$ varies from $\pi/6$ to zero, and $\chi$ varies from zero to $\pi/6$, when $\theta_s$ varies from $\pi/2$ to zero.
The degree of polarization will simply determine by using equation \ref{deg-polz-un}, and in general, for a given scattered meridian $(\phi_i+\phi_{is})$-plane  the P varies from $P(\theta=\phi_{is})$ of equation (\ref{deg-polz-un}) to zero for the variation of $\theta_s$ from $\pi/2$ to zero respectively. 
The results for $P$ as a function of $\theta_s$ for 6 different meridian planes with $\phi_{is}$ = 15, 30, 45, 60, 75 and 90 degree are shown in table \ref{tab-I-II}.

\subsection{Case III: any fixed incident photon direction and a random scattering plane}
Next, we fix the $k$ direction ($\theta_i,\phi_i$) and take all possible directions of ($k \times k'$) 
on a perpendicular plane to $k$. Like case II, the incident photon is scattered in all directions. For a given meridian plane,  each $k'$ is arisen from a different 
direction of ($k \times k'$), and clearly the different $\theta$. The PD can be determined using equation (\ref{deg-polz-un}) for known $\theta$. 
Since, the scattering plane  is unique for a given $k'$, like PD, PA 
will be also unique. We have computed the $P$ and $\chi$ for the meridian plane ranges from $\phi_i$ to $\phi_i +\pi$, and for each meridian plane $\theta_i$ ranges from 0 to $\pi/2$.
It should be noted that case III is identical to case I  for $\theta_i$ = 0. 

\subsection{Case IV: $k$ lies on the surface of cone at centre with opening angle $\theta_i$ and a random scattering plane}
We extend the case III with considering that $\theta_i$ of $k$ is still fix but now $\phi_i$ can take any value in [0,2$\pi$], so essentially the vector $k$ is rotating on the surface of the cone of opening angle $\theta_i$ which is situated at the centre.
Unlike Case II or III, here on a given meridian plane the $k'$ is arisen by
any  $k$ with appropriate scattering angle, thus PD and PA of $k'$ would be estimated by averaging method.
The averaged $P$ and $\chi$ of the $k'$ are computed using equation (\ref{avg_dpa}) for a given $\theta_s$, 
where now, n is total no. of scattered photons with $\theta_s$, 
the $p_j$ is determined for each scattered photon for known $\theta$ by using equation (\ref{deg-polz-un}).  
Since, for unpolarized incident photons the $k'_e$ lies 
  along the $(k\times k')$ which gives $\phi_j$ = 0, but we have fixed the reference $(k\times k')\text{-next}$ to measure the PA, 
so $\phi_j$ is  the angle between $(k\times k')$ and meridian plane ($(k\times k')$-next).
Expectedly, the variation of $P$, and $\chi$ as a function of $\theta_s$ is same for all scattered meridian planes.  In table \ref{tab-IV}, we have noted the results for $\theta_i$ = 0, 15, 30, 45, 60, 75 and 90 degree along with $f_{\theta_s}$ (a total no. of scattered emergent photons with $\theta_s$ on a given meridian plane).
For example, the emergent photons along the $z$-axis ($\theta_s$=0) are unpolarized for any opening angle $\theta_i$, as in this case for all incident photons the scattering angle is $\theta = \theta_i$, and the electric vector of $k'$ is isotropically distributed.
In general, for a given $\theta_i$, the emergent photons escape maximally at %
$\theta_s$ $\sim$ $\theta_i$,  
e.g., for $\theta_i$ = 15$\degree$ the maximum emergent photons escape with $\theta_s$ $\sim$25$\degree$ and have PD $\sim$ 0.06.

We compare the above results with the results of case III. Since 
the given $k'$ can be arisen from the incident meridian ($\phi_s+\phi_{is}$)-plane with $\phi_{is}$ $\equiv$ [-$\pi$, $\pi$] of case III, or 
particularly $n$ different situations of case III in range of $\phi_i$ $\equiv$ $[0,2\pi]$. As %in case III,
the $P$ and $\chi$ of case III are uniquely determined 
for a given $\phi_{is}$, the resultant $P$ and $\chi$ of case IV for a given $\theta_s$  can be obtained by weighted  averaging method, as defined in equation (\ref{avg_dpa}). For present case this equation is rewritten as   
\beqn \label{avg_dpa1} P \cos(2(\xi+\chi)) + 1 = \sum_{j=1}^{n} f_j \left [p_j\cos(2(\xi+\chi_j))+1\right], \eeqn
here, $f_{j}$ is the fraction of emergent scattered photons at $\theta_s$ to total that arisen due to the $j^{th}$ incident meridian  $(\phi_s+\phi_{is})$-plane, and $p_j$, and $\chi_j$ are corresponding PD and PA respectively. 
We find that the results are  similar in both ways.
\begin{table*}%[h!]\vspace{-0.0cm}
%%\captionsetup{font=footnotesize}
\centering
\caption{The PD and PA of the emergent photons for case IV along with $f_{\theta_s}$  (total no. of the emergent photons for a given $\theta_s$). The results are presented for seven different opening angles of cone (on which incident photon lies), $\theta_i$ = 0, 15, 30, 45, 60, 75 and 90 degree. }
\begin{scriptsize}
\begin{tabular}{|p{0.7cm}*{7}{p{1.74cm}}| }
  \hline
$\cos(\theta_s)$   & \multicolumn{7}{|c|}{$P \ \ (\chi, \ f_{\theta_S} )$ for case IV when $\theta_i$ = (in degree)} \\ \cline{2-8}
  & \multicolumn{1}{|l}{0}& 15 &  30 & 45  & 60 & 75 & 90 \\ \hline
  0.0 & 0.996 (90.9, 1.73) & 0.866 (90.9, 1.87) & 0.558 (89.0, 2.29) & 0.220 (89.0, 3.13) & 0.014 (180., 4.25) & 0.132 (180., 6.71) & 0.078 (180., 16.1)\\
  0.1 & 0.969 (90.9, 1.76) & 0.842 (89.0, 1.86) & 0.536 (90.9, 2.37) & 0.213 (89.0, 3.12) & 0.020 (165., 4.48) & 0.120 (01.8, 6.99) & 0.124 (180., 9.75)\\
  0.2 & 0.908 (90.9, 1.91) & 0.783 (89.0, 2.06) & 0.491 (89.0, 2.54) & 0.180 (90.9, 3.18) & 0.031 (07.2, 4.73) & 0,100 (180., 8.18) & 0.150 (180., 7.17)\\
  0.3 & 0.816 (90.9, 2.04) & 0.703 (89.0, 2.17) & 0.431 (89.0, 2.61) & 0.152 (87.2, 3.55) & 0.040 (176., 5.12) & 0.089 (01.8, 8.49) & 0.155 (180., 6.03)\\
  0.4 & 0.703 (90.9, 2.30) & 0.596 (90.9, 2.46) & 0.351 (89.0, 2.92) & 0.112 (90.9, 3.81) & 0.034 (05.4, 6.09) & 0.114 (180., 5.75) & 0.157 (180., 4.88)\\
  0.5 & 0.581 (90.9, 2.59) & 0.491 (89.0, 2.78) & 0.275 (90.9, 3.23) & 0.068 (89.0, 4.38) & 0.026 (03.6, 9.32) & 0.116 (180., 4.35) & 0.150 (178., 3.98)\\
  0.6 & 0.454 (90.9, 3.08) & 0.374 (90.9, 3.18) & 0.193 (90.9, 3.67) & 0.027 (87.2, 5.31) & 0.051 (05.4, 5.06) & 0.100 (03.6, 3.39) & 0.122 (180., 3.07)\\
  0.7 & 0.328 (90.9, 3.50) & 0.269 (89.0, 3.85) & 0.125 (89.0, 4.40) & 0.004 (69.0, 8.83) & 0.056 (05.4, 3.61) & 0.094 (180., 2.57) & 0.109 (178., 2.22)\\
  0.8 & 0.208 (90.9, 4.01) & 0.160 (90.9, 4.24) & 0.057 (90.9, 5.65) & 0.006 (34.5, 4.43) & 0.054 (174., 2.60) & 0.075 (176., 1.90) & 0.084 (03.6, 1.59)\\
  0.9 & 0.098 (90.9, 4.63) & 0.066 (89.0, 5.17) & 0.017 (83.6, 5.64) & 0.016 (169., 2.65) & 0.035 (180., 1.64) & 0.041 (01.8, 1.10) & 0.035 (09.0, 0.98)\\
  1.0 & 0.000 (90.9, 2.62) & 0.003 (96.3, 0.16) & 0.012 (170., .09)* & 0.044 (101., .05)* & 0.024 (76.3, .02)* & 0.188 (112., .01)* & 0.301 (158., .01)*\\
  \hline
\multicolumn{8}{l}{* the corresponding value is not reliable due to low statistics. } \\  
\multicolumn{8}{l}{Note: for $\theta_i = \theta_s$, one has small $P$ ( in range of (0.002-0.08)) and maximum $f_{\theta_s}$.}\\
\end{tabular}\vspace{0.3cm}\end{scriptsize}
\label{tab-IV}
\end{table*}

\subsection{Case V: general case}

Finally, we consider a general case where there is no restriction on $k$. That is, the scattered photon $k'$ is arising from all directions of $k$. We compute the $P$ as a function of $\theta_s$ for any meridian $\phi_s$-plane, 
We find $P$ $\approx$ 11, 0 $\%$ for $\theta_s$ = 90, 0 degree respectively.
The $\chi$ is $\sim$ 90 degree for all $\theta_s$. Since we are measuring $\chi$ with respect to $(k\times k')$-next, 
it signifies that $k'_e$ is parallel to the $(x,y)$-plane.
Clearly, the all scattering planes (which generate the $k'$ on $\phi_s$-plane) are not a $\phi_s$-plane, but
we notice that on averaged the scattering plane is mostly a scattered meridian plane, thus the results $\chi$ $\sim$ 90$\degree$ confirms that the polarization of single scattered unpolarized (incident) photons is perpendicular to the scattering plane. The results are shown in table \ref{tab-gen-res}.

Like case IV, the results of case V are verified by a weighted averaging method using the result of case IV. For convenience in table \ref{tab-IV} we have also listed the $f_{\theta_s}$  along with the $P$ and $\chi$. Thus, $f_j = \frac{f_{\theta_s}}{\Sigma_{j=1}^n f_{\theta_s}}$ for a given $\theta_s$ and $\theta_i$ (here, $i=j$), and with having $n = 7$ we obtain the results of case V approximately using table \ref{tab-IV} and equation (\ref{avg_dpa1}).

The results are qualitatively agreed with almost century old calculations of \cite{Chandrasekhar1946} \cite[see also,][]{Chandrasekhar1960}.
\cite{Chandrasekhar1946} had solved the radiative transfer equations, which is governed by the Thomson scattering by free electron, for the intensities of two states of polarization, one $I_l$ is along the meridian plane and other one $I_r$ is perpendicular to it, with no incident radiation. In defining the source function for $I_l$ or $I_r$, the considered cross section for the %direction of
polarization in perpendicular to and parallel to the scattering plane is either for unpolarized photons, or combinations of two polarized photons with polarization angle $\phi$ and $\phi + \pi/2$, see equations (2) and (3) of their paper and table 2 for the results (and for refined results see table XXIV in \citealp{Chandrasekhar1960}). So, the laws of darkening for the PD of the emergent photon is only for the single scattering event in Thomson regime for unpolarized incident photons. 
In the next section, we compute the laws of darkening for multi scattering events.

\begin{table*}%[h!]\vspace{-0.0cm}
%%\captionsetup{font=footnotesize}
\centering
\caption{The PD of the emergent photons for general case V from any given meridian plane after single scattering of the randomly oriented unpolarized incident photons 
(or, the re-estimation of laws of darkening of \citet{Chandrasekhar1946} with having a general Klein-Nishina cross section, equation \ref{gen-dif-polz} ).
}\vspace{-0.3cm}
\begin{footnotesize}
  \begin{tabular}{|p{0.80cm}*{11}{p{0.8cm}}| }
    \hline 
  $\cos(\theta_s)$ & 0.0 & 0.1 & 0.2 & 0.3 & 0.4 & 0.5 & 0.6 & 0.7 & 0.8 & 0.9 & 1.0 \\
  $P$ & 0.111 & 0.112 & 0.103 & 0.097 & 0.097 & 0.082 & 0.071 & 0.059 & 0.042 & 0.020 & 0.010 \\
  \hline
\end{tabular}\vspace{0.3cm}\end{footnotesize}
\label{tab-gen-res}
\end{table*}

For all cases, we have estimated the PD and PA by two another methods, 
(i)  we
perform second scattering and construct the modulation curve for either $\theta = 0$ or $\theta \equiv [0,25\degree]$ to estimate the $P$ as discussed in section \S\ref{theta_0}, (ii) we construct the modulation curve using the probability $p(\eta)$ (of equation \ref{gen-modu-p}) and estimate the $P$, as discussed in section \S\ref{sec:gen-modu-p}. We find that in all
cases the results match qualitatively with these two methods.

\section{General results and Comparison with observations}\label{sec:multi}

In previous section we estimate the polarization for a single scattering event
in Thomson regime and for a almost rest electron. In this section we consider %
a multi scattering event with arbitrary electron Lorentz factor. We explore the general results for spectro-polarimatric measurement for XRBs source, and make a comparision for few XRBs observed by IXPE. 
In general, a detailed accretion disk + corona geometry is required for the  spectro-polarimatric study for XRBs \cite[e.g.,][]{Beheshtipour-etal2017, Schnittman-Krolik2010}.
However, here to estimate the polarization for multi scattering events and the energy dependency of polarization, for completeness %of previous finding,
we consider the same spherical corona geometry (as described in section \S\ref{mc:method}). %
Therefore our results will describe the observed polarization properties qualitatively only.  

\subsection{Multi scattering}
To verify the MC calculations for multi scattering we consider the case I with $\theta_i \equiv [0,\pi]$, as in this case the scattered photon has PD $\sim$0.33 in all directions of $\theta_s$.
We first compute (say, first method) the $P$ value of 2nd times scattered photon for this case.
Next, we consider (say, 2nd method) a partially polarized incident photon with $P$ = 0.33, $\chi$ = 90 degree and $(k \times k')$ lies either on the meridian plane or perpendicular to the meridian plane (and here, for 0.66 fraction of unpolarized photons, the $(k \times k')$ lies randomly), and estimate the $P$ after single scattering. We find that for both methods the result qualitatively agrees, with $P$ $\sim 0.1-0.12$ on a given meridian plane.

Interestingly, we find that for all cases, I $-$ V, after 3-4 numbers of scattering, the maximum $P$ value is reduced to $\sim$ 0.02 - 0.05 on any meridian plane. Here, again we consider a Thomson regime and almost rest electron. Therefore, for all cases and average scattering number $>$ 4, the emergent scattered photon is mainly unpolarized with averaged maximum $P \sim 0.035$. Here, we like to point out that if these emergent photons again freshly scatter in optically thin corona with average scattering number $\sim$1 then the $P$  $\&$  $\chi$ will be described like case V.

\subsection{Energy dependency of polarization}
The prime focus is here to study the polarization properties for X-ray binaries,
XRBs frequently transit from soft spectral state to hard state and vice versa. To understand and to explore the energy dependency of polarization for Comptonized photons, we consider two different steady spectral states with unpolarized seed photons. The first is a soft spectrum (Model 1) with low electron medium temperature $kT_e$ $= 2.5$ keV \cite[see, e.g.][]{Kumar-Misra2014}. In view of neutron star NS low-mass X-ray binaries, we consider two different seed photon source (black body) temperatures corresponding to Hot-seed and Cold-seed photon model \cite[see, e.g.][references therein]{Kumar-Misra2016a} which are $kT_b$ $=$ 1.5 and 0.7 keV respectively. We refer Model 1a with temperature $kT_e =$ 2.5keV and $kT_b =$ 1.5 keV and for Model 1b $kT_e =$ 2.5keV and $kT_b =$ 0.7 keV.
Since, for consistency we consider only a spherical corona geometry, but these two models are defined based on the seed photon source geometry \cite[e.g.,][]{Lin-etal2007}, thus our study does not give physical insight of the model but only provides the dependency of PD and PA  on $kT_b$ variations.
The second is a hard spectrum (Model 2) with high electron medium temperature $kT_e =$ 100 keV and $kT_b =$  0.3 keV.
To explore the general results we take two optical depth values in such a way that the corresponding average scattering number $\langle N_{sc}\rangle$ $\approx$ 1.1 and 5.0.
Thus, conclusively to explore the energy dependency of PD and PA, we take mainly three Models 1a, 1b and 2 with two values of $\langle N_{sc}\rangle$ $\approx$ 1.1 and 5.0.
PD and PA have been computed for four values of $\theta_s$ = 30$\degree$, 45$\degree$, 60$\degree$, 75$\degree$. In considered spherical geometry, if one assumes, the $z$-axis is along the radio-jet direction, then the $(x,y)$-plane will mimic the accretion disk and so the angle $\theta_s$ is equivalent to the disk inclination angle $i$ (or the angle between the line of sight and the normal to the disk plane).
Thus we also study the variation of PD and PA with disk inclination angle $i$.
And now, in present convention, PA = 90$\degree$ signifies that the electric vector is parallel to the disk plane. %

\begin{figure*}%[h!]\vspace{-0.19cm}
%%\captionsetup{font=footnotesize}
\centering
\begin{tabular}{lll}%\hspace{-1.5cm}
  \includegraphics[width=0.37\textwidth]{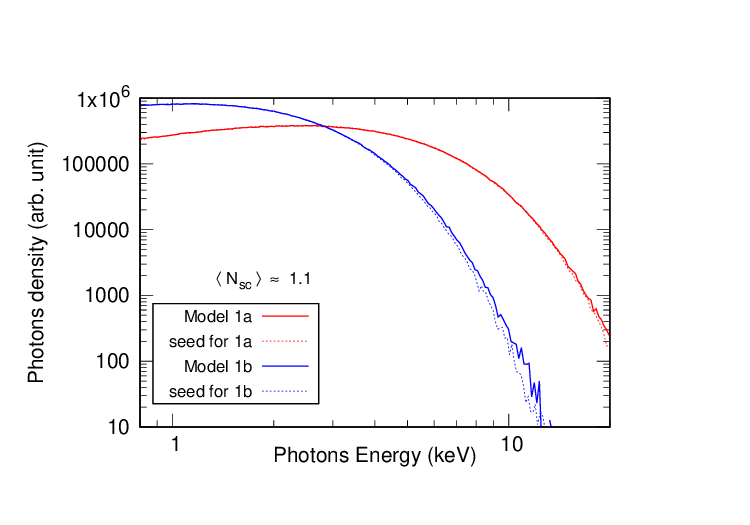} &\hspace{-1.5cm}%sed_mod1_1n.eps
    \includegraphics[width=0.37\textwidth]{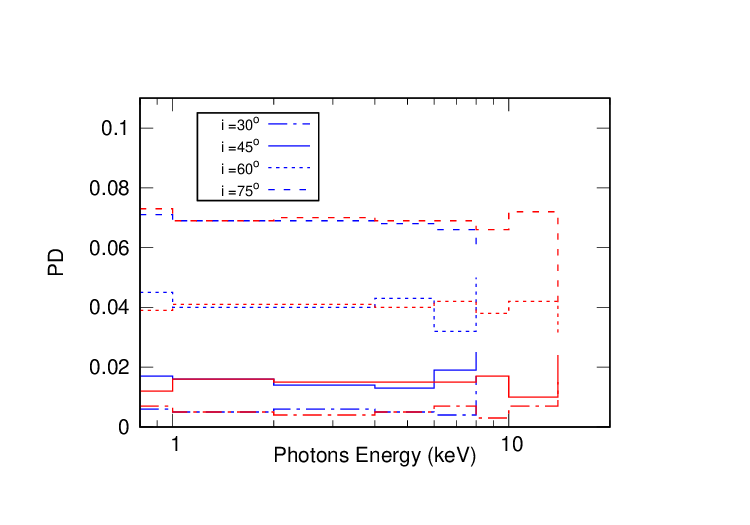} &\hspace{-1.5cm}%dp_mod1_1gsn.eps
    \includegraphics[width=0.37\textwidth]{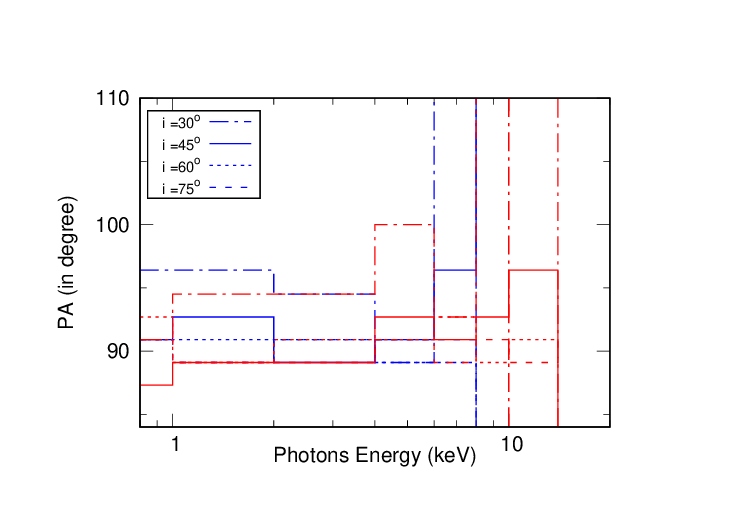}\vspace{-1.0cm}  \\%da_mod1_1gsn.eps\hspace{-1.5cm}
  \includegraphics[width=0.37\textwidth]{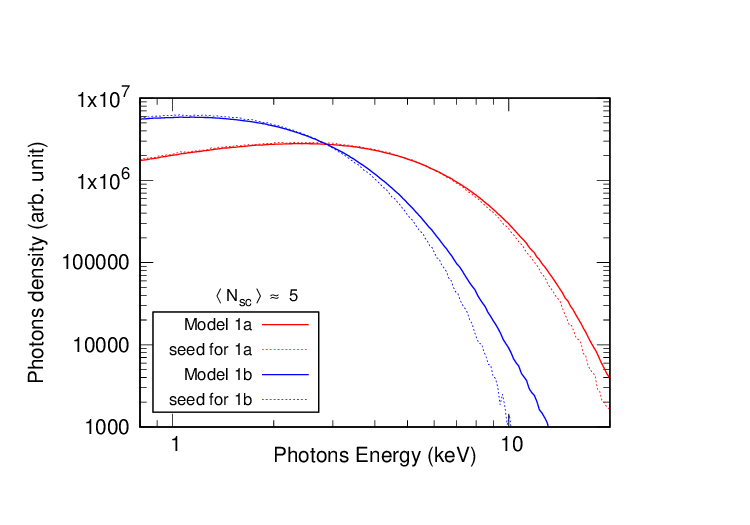} &\hspace{-1.5cm}%sed_mod1_5n.eps
    \includegraphics[width=0.37\textwidth]{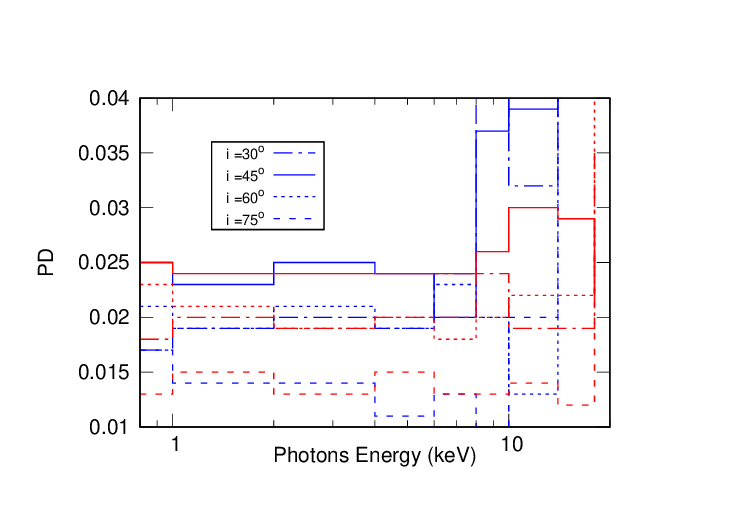} &\hspace{-1.5cm}%dp_mod1_5n.eps
    \includegraphics[width=0.37\textwidth]{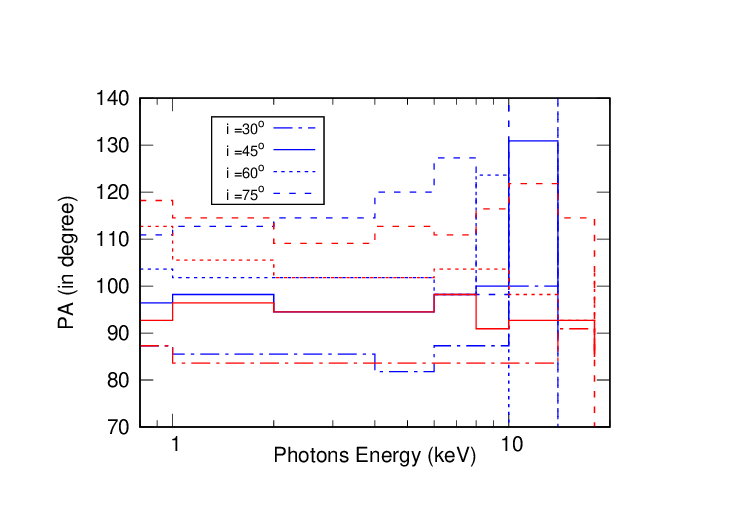}  \\%da_mod1_5n.eps
  
\end{tabular}\vspace{-0.3cm}
\caption{Comptonized photons density distribution, degree of polarization and angle of polarization are shown in left, middle and right panels respectively when the seed photons (shown by dashed line in left panel) are unpolarized. The upper and lower panels are for the average scattering number $\approx$ 1.1 and 5 respectively. PD and PA are computed for four $\theta_s$ (or, alternatively the disk inclination angle $i$) values = 30, 45, 60 and 75 degree, which are shown by dashed-dotted, solid, dotted and dashed lines respectively. PA is computed on perpendicular plane to the escaped photon directions, not on the sky plane, here PA = 90$\degree$ signifies that the electric vector parallel to the ($x,y$)-plane or accretion disk, see the text for details. The energy bins are 0.8, 1, 2, 4, 6, 8, 10, 14, 18, 22, 26 and 30 keV.}
\label{fig:mod1}
\end{figure*} 

The general results for Model 1a and 1b are shown in Figure \ref{fig:mod1}. 
The upper and lower panels are for $\langle N_{sc}\rangle$ $\approx$ 1.1 and 5 respectively. In the left panel, the seed photon flux is shown by dashed curve and Comptonized photon flux by solid line. As expected for $\langle N_{sc}\rangle$ $\approx$ 1.1 the PD as a function of $\theta_s$ (or, $i$) is slightly lower than the respective values listed in table \ref{tab-gen-res}. The PD values for given $i$ are almost constant over the energy bin, also qualitatively independent from the seed photon source temperature $kT_b$.
Due to the low photons statistics, the PD and PA are computed upto photon energy 8 and 16 keV for Model 1a and 1b respectively, and we also notice that %
their values are fluctuated around this energy bin. Since for unpolarized incident photons and for single scattering, it is expected that the electric vector of Comptonized photons will lie normal to the scattering plane. And for $\langle N_{sc}\rangle$ $\approx$ 1.1 we find almost the same, here for all $i$ values, the PA lies from 89$\degree$ to 95$\degree$.

\begin{figure}%[h!]\vspace{-0.19cm}
%%\captionsetup{font=footnotesize}
\centering
\begin{tabular}{l}%\hspace{-1.5cm}
  \includegraphics[width=0.37\textwidth]{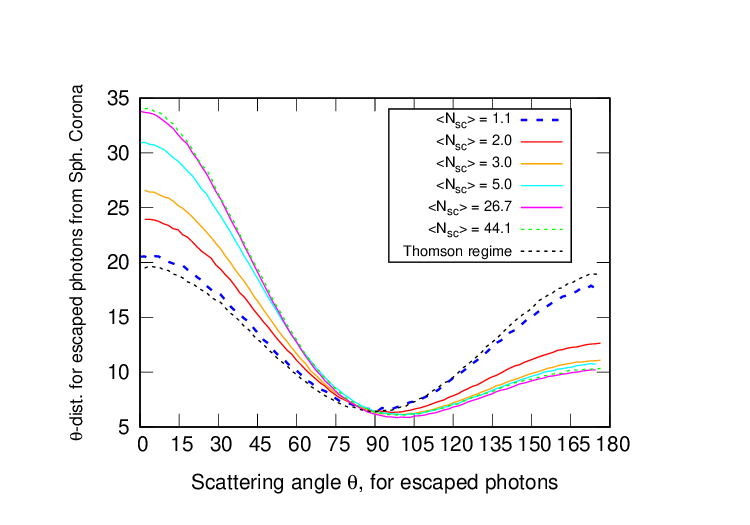}%theta_dist-sph-corona.eps   
\end{tabular}\vspace{-0.3cm}
\caption{The $\theta$-angle distribution of escaped Comptonized photons in spherical corona after experienced of average scattering number $\langle N_{sc}\rangle$. The blue, red, orange, cyan, magenta and green curves are for $\langle N_{sc}\rangle$ = 1.1, 2.0, 3.0, 5.0, 26.7 and 44.1 respectively. The Klein-Nishina cross section in Thomson regime is shown by black dotted curve. Note, the Comptonized photons which are inside the medium always follows Klein-Nishina cross section for any given $\langle N_{sc}\rangle$. }
\label{fig:escap-theta}
\end{figure}

In model 1, the spectral parameters are in Thomson regime, so the distribution of scattering angle will be %follow
$\propto (1+ \cos^2\theta)$. In considered spherical corona geometry, the Comptonized photons after any number of scattering follow the same scattering angle distribution of Thomson regime. But the Comptonized photons which escape the medium (corona) do not follow it for $\langle N_{sc}\rangle >$ $\sim$ 1.1. The deviation from Thomson cross section for escaped Comptonized photons can be understood as. Supposed, before escaping the medium the scattered photon is at scattering site whose  distance is 0.7L from the centre, then to escape the medium in forward direction the photon has to travel collision  free path of length 0.3L but in case of backward direction the collision free path length is 1.7L. With having the exponential distribution for the collision free path it is more likely, on average (as the probability for travelling the scattered photon in forward or backward direction is same), that the photon will escape the medium in a forward direction in comparison to the backward direction. In addition, the trend for escaping the photons in forward direction increases with $\langle N_{sc}\rangle$, as the mean free path for photons decreases with increasing optical depth.
For a given medium, it is also expected that after some value of  $\langle N_{sc}\rangle$ the scattering angle distribution will get saturated, we find, the saturation occurs around $\langle N_{sc}\rangle$ = 25. The results are shown in Figure \ref{fig:escap-theta}.

\begin{figure*}%[h!]\vspace{-0.19cm}
%%\captionsetup{font=footnotesize}
\centering
\begin{tabular}{lll}%\hspace{-1.5cm}
  \includegraphics[width=0.37\textwidth]{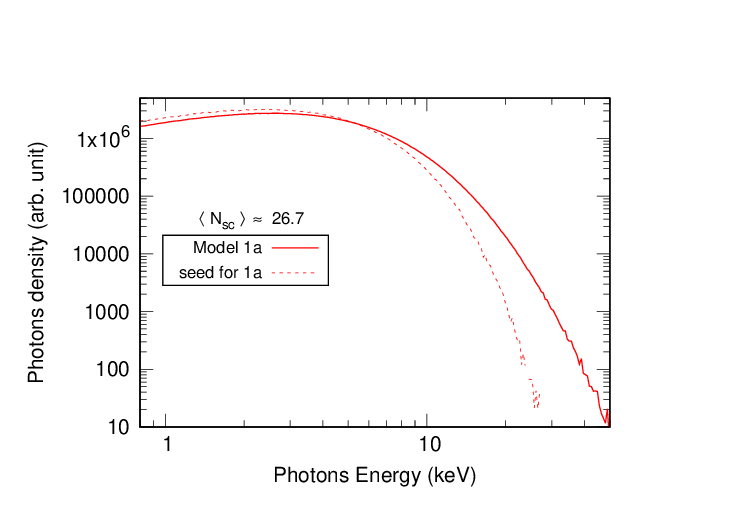} &\hspace{-1.5cm}%sed_mod1a_26n.eps
    \includegraphics[width=0.37\textwidth]{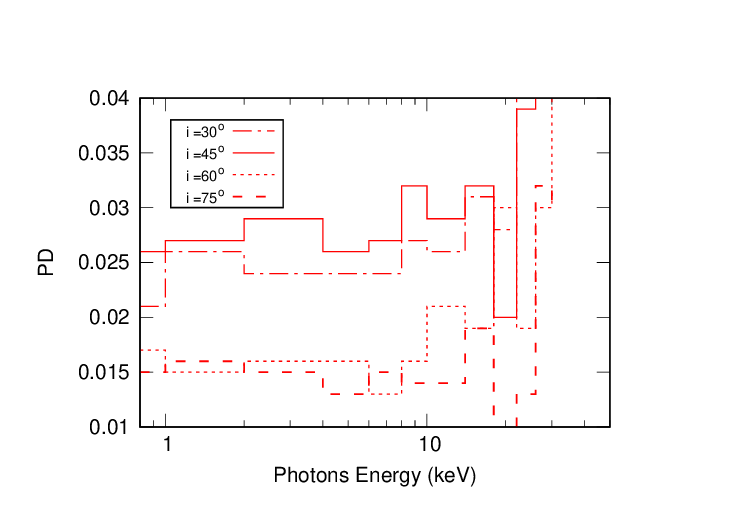} &\hspace{-1.5cm}%dp_mod1a_26n.eps
    \includegraphics[width=0.37\textwidth]{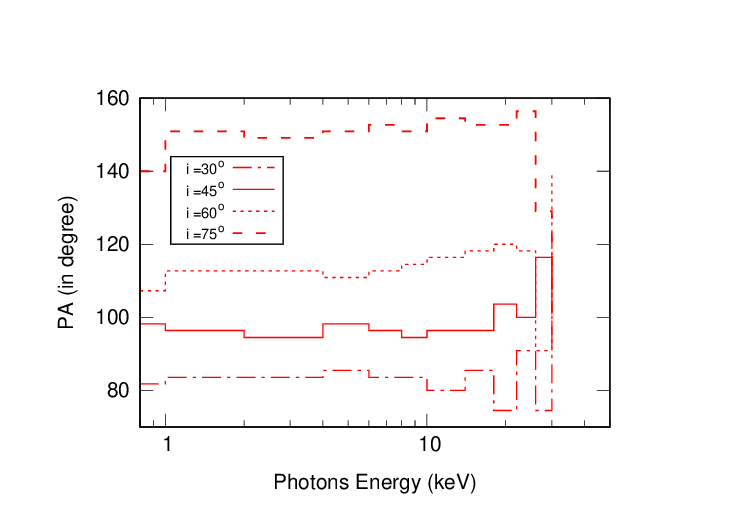} \\%da_mod1a_26n.eps
\end{tabular}\vspace{-0.3cm}
\caption{For spectral parameter Model 1a and the average scattering number $\approx$ 26.7. The rests are same as Figure \ref{fig:mod1}.  }
\label{fig:mod1_26}
\end{figure*}

For $\langle N_{sc}\rangle$ $\approx$ 5 (in lower panel of Figure \ref{fig:mod1}), the PD and PA are calculated upto photon energy 14 and 22 keV for Model 1b and 1a respectively. Like, $\langle N_{sc}\rangle$ $\approx$ 1.1, the PD and PA are independent of the k$T_b$. PD is almost constant over the photon energies ($<$ 10 keV). The fluctuation in PD and PA values above 10 keV is due to the low photons statistics. PA values range from 80$\degree$ to 120$\degree$ when the $i$ ranges from 30$\degree$ to 75$\degree$. We observe that the $\theta$-angle distribution for escaped Comptonized photons
for a given $i$ has an extra small hump (by a factor $\sim$ 1.1 $-$ 1.2 from the corresponding values of cyan curve of Figure \ref{fig:escap-theta}) around $\theta = i$, which leads to a maximum value of PD for $i$ = 45$\degree$. %
Here, we find the  maximum value of PD $\sim 0.025$ for $i$ =45$\degree$.
In general, in the soft state of XRBs the optical depth is relatively high, to characterize this we consider Model 1a with $\langle N_{sc}\rangle$ $\approx$ 26.7. %
The results are shown in Figure \ref{fig:mod1_26}. Here, the magnitude of PD values is similar to the case of $\langle N_{sc}\rangle$ $\approx$ 5, only the dependency of PD value on $i$ has changed. The range of PA values is wider now and for $i$ = 75$\degree$ PA is $\sim$ 150$\degree$. 


\begin{figure*}%[h!]\vspace{-0.19cm}
%%\captionsetup{font=footnotesize}
\centering
\begin{tabular}{lll}%\hspace{-1.5cm}
  \includegraphics[width=0.37\textwidth]{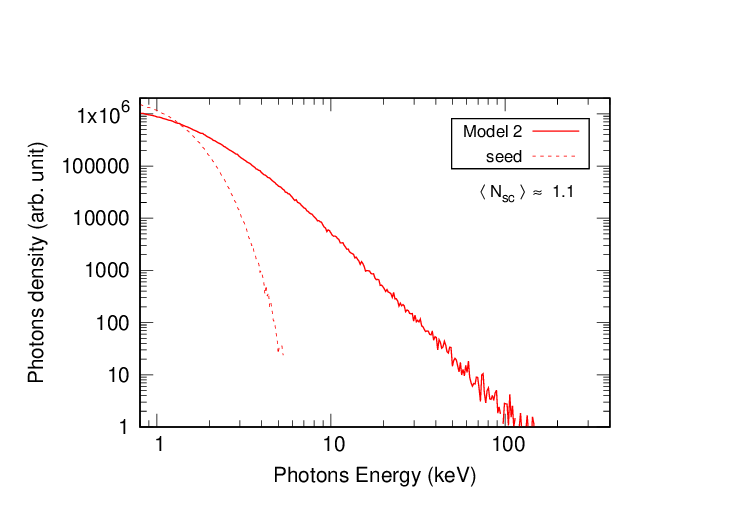} &\hspace{-1.5cm}%sed_mod2_1n.eps
    \includegraphics[width=0.37\textwidth]{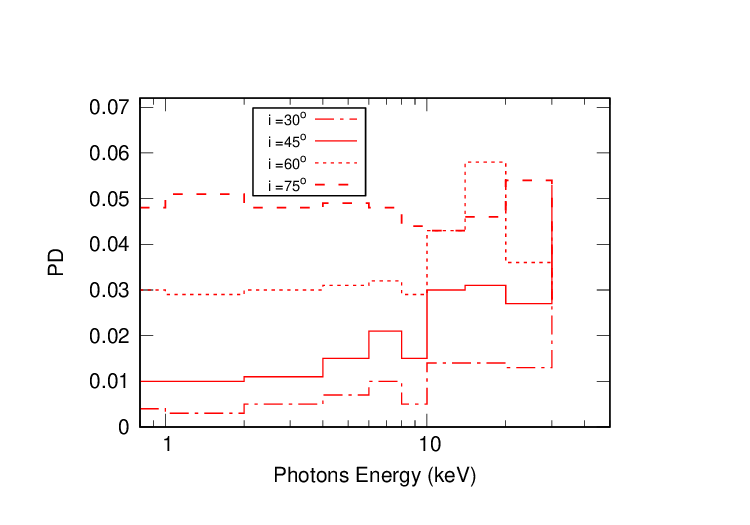} &\hspace{-1.5cm}%dp_mod2_1gsn.eps
    \includegraphics[width=0.37\textwidth]{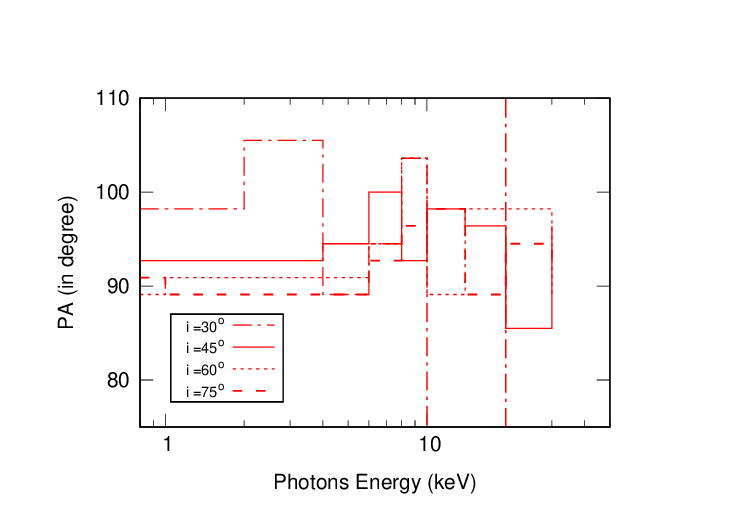}\vspace{-1.0cm}  \\%da_mod2_1gsn.eps\hspace{-1.5cm}
  \includegraphics[width=0.37\textwidth]{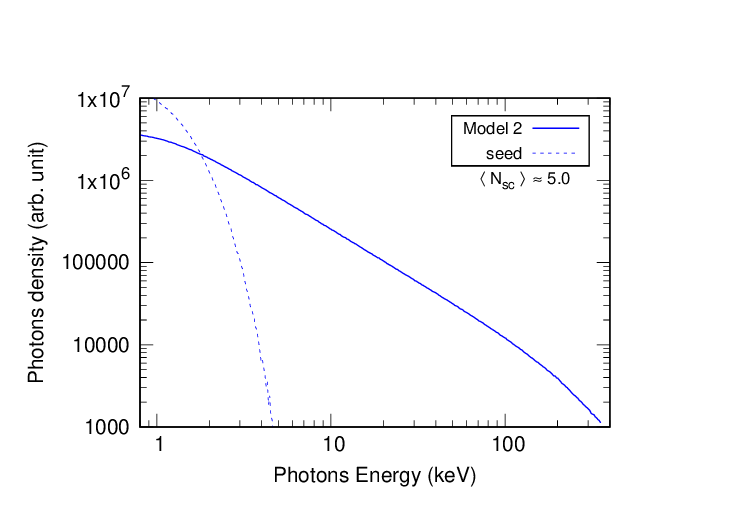} &\hspace{-1.5cm}%sed_mod2_5n.eps
    \includegraphics[width=0.37\textwidth]{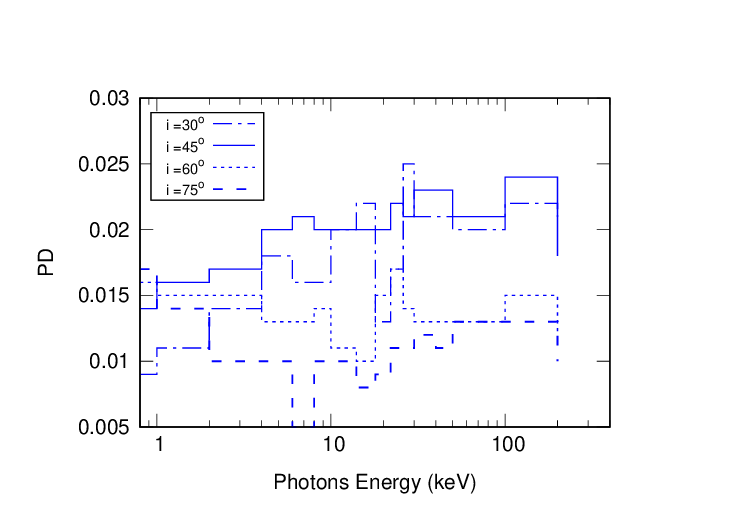} &\hspace{-1.5cm}%dp_mod2_5n.eps
    \includegraphics[width=0.37\textwidth]{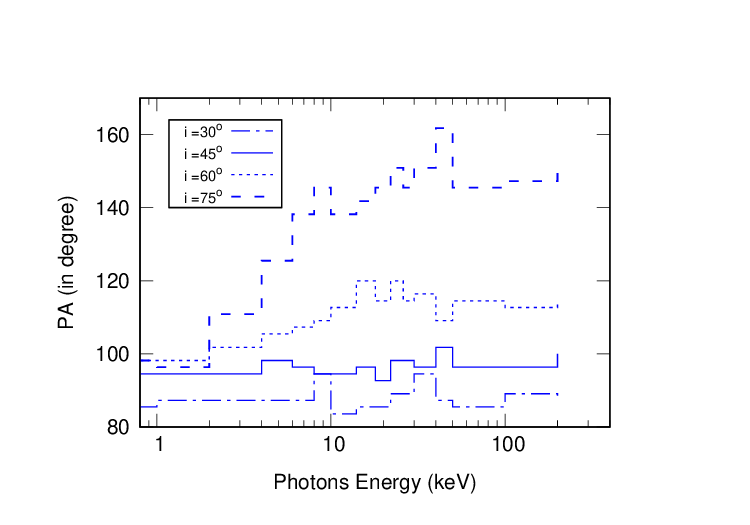}  \\%da_mod2_5n.eps
  
\end{tabular}\vspace{-0.3cm}
\caption{For spectral parameter Model 2 and the upper and lower panels are for average scattering number $\approx$ 1.1 and 5.0 respectively. The energy bins are 0.8, 1, 2, 4, 6, 8, 10, 14, 18, 22, 26, 30, 40, 50, 100 and 200 keV. The rests are same as Figure \ref{fig:mod1}.}
\label{fig:mod2}
\end{figure*}

In Figure \ref{fig:mod2}, the general results for Model 2 have been shown. The upper and lower panels are $\langle N_{SC}\rangle$ $\approx$ 1.1 and 5. For $\langle N_{SC}\rangle$ $\approx$ 1.1, the PD and PA have been computed upto 30 keV. Like Model 1, PD is constant over photons energy ($<$ 10keV), and PA is $\sim$ 90$\degree$ and independent to the photons energy ($<$ 6keV). The magnitude of PD is slightly lower in comparison to the Model 1, e.g., for $i$ = 75$\degree$ PD = 0.07 and 0.05 for Model 1 and 2 respectively. Also, for above 6 keV photons energy the PA ranges 95$\degree$ - $\sim$100$\degree$. Here, the decrement in PD values and the deviation of PA from 90$\degree$ is mainly due to the multi scattering. Since, in Model 2 due to the large $kT_e$ = 100 keV and low $kT_b$ = 0.3keV, the photons of energy $>$ 2keV on average experienced a large scattering number from the averaged value 1.1. 
For example, for 2 $-$ 10 keV photons the averaged scattering number varies from 1.1 to 1.5, while for 10 $-$ 70 keV it varies from 1.5 to 2.5.  
For $\langle N_{SC}\rangle$ $\approx$ 5, as expected the PD value is slightly lower in comparison to the corresponding value of Model 1. The increasing
behaviour %characteristics
of PD for low photons energy ($<$ 10 keV) is mainly corresponded to the $\theta-$angle distribution for escaped photons, as the photons with energy less than 7 keV have average scattering number less than 5. The PA values range from 80$\degree$ to 150$\degree$.

\begin{figure*}%[h!]\vspace{-0.19cm}
%%\captionsetup{font=footnotesize}
\centering
\begin{tabular}{lll}%\hspace{-1.5cm}
  \includegraphics[width=0.37\textwidth]{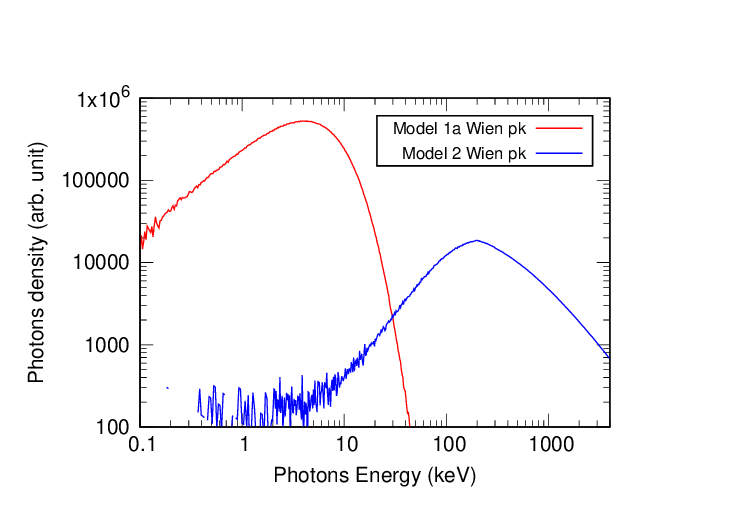} &\hspace{-1.5cm}%sed_mod12wien.eps
    \includegraphics[width=0.37\textwidth]{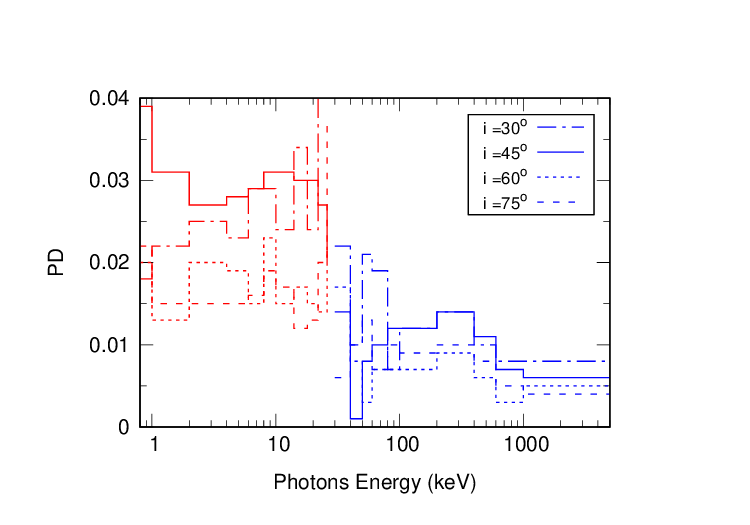} &\hspace{-1.5cm}%dp_mod12wien.eps
    \includegraphics[width=0.37\textwidth]{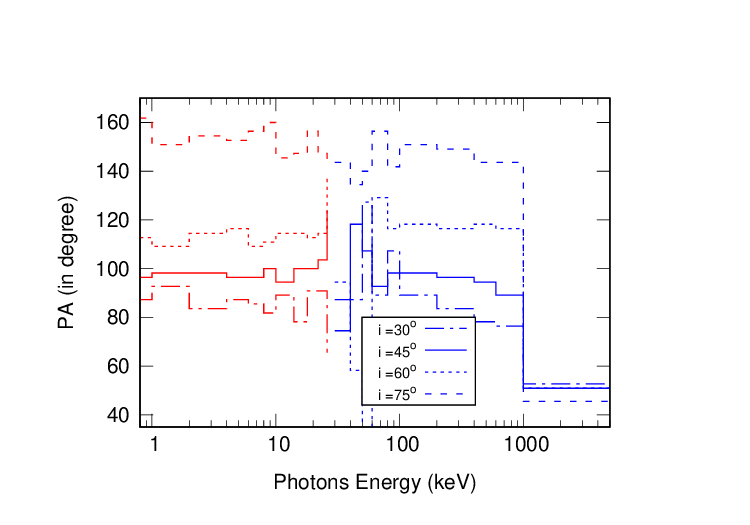} \\%da_mod12wien.eps
\end{tabular}\vspace{-0.3cm}
\caption{For Wien spectra 
  of Model 1a and 2 with average scattering number $\approx$ 170 and 45 respectively. The energy bins for Model 1a are 0.8, 1, 2, 4, 6, 8, 10, 14, 18, 22 and 26 keV and for Model 2 are 30, 40, 50, 60, 80, 100, 200, 400, 600, 1000 and 5000 keV. The rests are same as Figure \ref{fig:mod1}.  }
\label{fig:mod12_wien}
\end{figure*}

For completeness, we have also computed the PD and PA for the Wien spectrum of Model 1 and 2, the results are shown in Figure \ref{fig:mod12_wien}. We know that the Wien spectrum does not depend on the seed photon spectrum but only depends on the
electron medium temperature. In addition, for low $kT_e$ (or non-relativistic electrons) one needs a large scattering number to generate the Wien spectrum, while for large $kT_e$ (relativistic electron) one need comparatively small scattering number \cite[e.g.,][]{Kumar-Kushwaha2021}.
We compute the Wien spectrum for Model 1 and 2 with $\langle N_{SC}\rangle$ = 170 and 45 respectively. For Model 1, the dependency of PD on $i$, and the range of PD are  similar to the case of Model 1a with $\langle N_{SC}\rangle$ = 26.7. While for Model 2 the magnitude of PD is around 0.01 upto photon energy 1000 keV.
Since, in the case of Model 2, in last, the seed photons are mainly $\sim$ 200 keV photons. From equation (\ref{deg-polz-un}) (or, see Figure \ref{fig:PD-unpolz})
we know that the PD value for 200 keV photons is comparatively smaller than the seed photon of energy 10 keV (or $<$ 10 keV), as a result we obtain a smaller value comparison to the Model 1.
Conclusively, The dependency of PD on $i$ of Wien spectra for unpolarized seed photons does not behave like case V, %,
see table \ref{tab-gen-res}. This also %asserts
indicates that the emergent photons from thin accretion disk 
(in which, the emergent spectrum is a black body due to the large optical depth by Thomson scattering \citealp{Shakura-Sunyaev1973})
would be  mainly unpolarized with maximum PD $\sim$ 0.03 for $i$ $\sim$ 45$\degree$. 

\begin{figure}%[h!]\vspace{-0.19cm}
%%\captionsetup{font=footnotesize}
\centering
\begin{tabular}{ll}\hspace{-1.5cm}
  \includegraphics[width=0.35\textwidth]{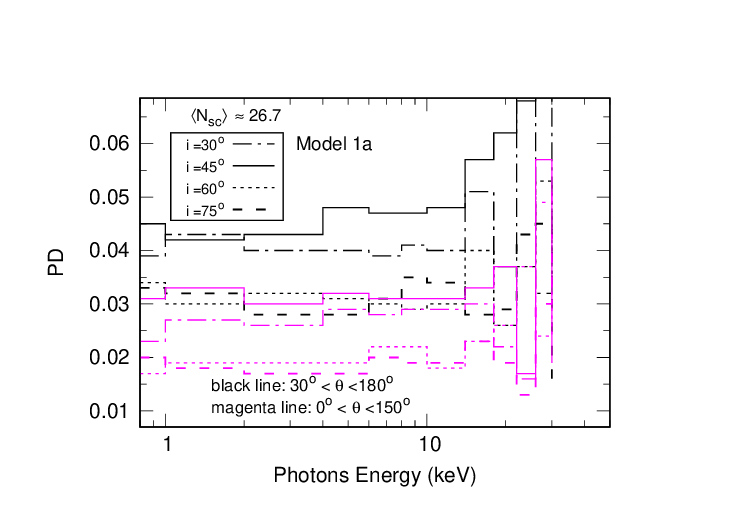} &\hspace{-1.5cm}%dp_mod1a_26con30n.eps
  \includegraphics[width=0.35\textwidth]{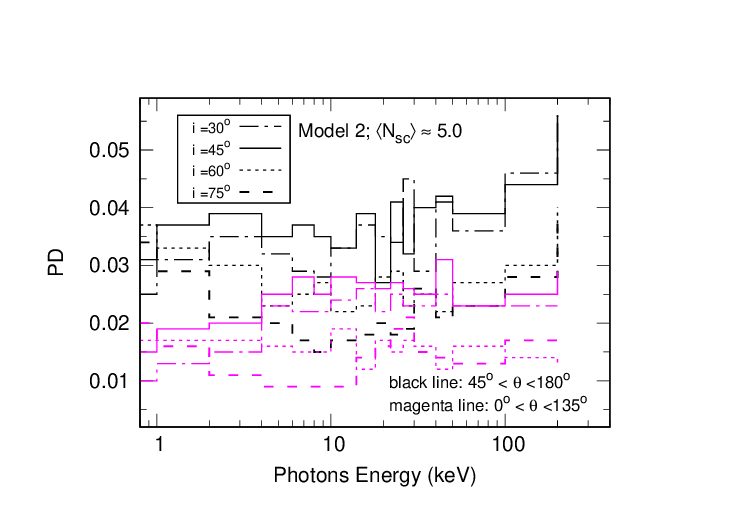}  \\%dp_mod2_5con45n.eps
\end{tabular}\vspace{-0.3cm}
\caption{ To understand the geometry affect in PD calculations, for simplicity we ad-hocly consider an asymmetric range of the scattering angle. The black and magenta curves are, in left panel  for $\theta$ $\equiv$ [30,180$\degree$] and [0,150$\degree$] and in right panel for $\theta$ $\equiv$ [45,180$\degree$] and [0,135$\degree$] respectively. The left and right panels are for Model 1a with $\langle N_{sc}\rangle\approx 26.7$ and Model 2 with $\langle N_{sc}\rangle\approx 5.0$ respectively. PD is computed for four $i$ values = 30, 45, 60 and 75 degree, and shown by dashed-dotted, solid, dotted and dashed lines respectively. For left panel the energy bins are same as Figure \ref{fig:mod1} while Figure \ref{fig:mod2} for right panel.}
\label{fig:mod1-2geo}
\end{figure} 

\subsection{Geometry dependency}

With having spherical corona we compute PD and PA for two extreme sets of spectral parameters. Particularly for large optical depth, the PD values are always less than 0.03 for photons of energy $<$ 10 keV. Recently IXPE has measured PD for many XRBs sources, and for a few sources the estimated PD is greater than the 0.03 in 2 $-$ 8 keV energy band \cite[e.g.][]{Krawczynski-etal2023, Jayasurya-etal2023}. Authors explained the comparatively large PD 
mainly in terms of different possible geometry for scattering medium in the disk.
In section \S\ref{sec:Chandra} we had noticed that the PD also depends on incident photon direction, see  the table \ref{tab-IV} for Case IV; also see, e.g., Case I for the variation of PD as a function of $\theta$.
Therefore, the dependency of PD on the geometry can be happened in terms of either distribution of incident photon direction or distribution of $\theta$-angle for the escaped Comptonized photons, or both.   
To understand the dependency of PD on geometry for simplicity we ad-hocly
exclude the some range of $\theta$ of the Comptonized escaped photons in estimation of PD without bothering about the spectra.
However, in appendix \ref{sp-pol} we argue that in Thomson limit the Comptonized spectrum is independent of $\theta$, and thus the spectra would not get changed in these cases.
The results are shown in Figure \ref{fig:mod1-2geo}. The left and right panels are for the Model 1a and 2 respectively. 
For Model 1a we consider two sets of $\theta$ range, [30,180$\degree$] and [0,150$\degree$]; while  [45,180$\degree$] and [0,135$\degree$] for Model 2. We find that PD values after excluding the backward direction are smaller than the PD obtained by excluding the forward direction for both models. This is because of the $\theta$-angle distribution of escaped photons, in which the maximum photons escape the medium with $\theta$ = 0, see Figure \ref{fig:escap-theta}. %Finally
Hence, for considered geometry for both models the magnitude of PD is around 0.04 or above when we exclude the forward direction of escaped photon. 

\subsection{Comparison with Observations}

In this section we briefly discuss the polarization properties of a few sources
observed by IXPE. 
However, these explanation
would be a purely qualitative, as we have a simple %istic
 spherical corona geometry, and also the present calculations have not been implemented in the general relativity formalism. Moreover, the main motive of this exercise is to what extent one can learn in understanding of polarization properties with having simple spherical corona geometry. We select five sources, in which two are black hole XRBs and three are neutron star XRBs.
\begin{itemize}
\item[] {\bf 4U 1630-47:} {\it IXPE} had observed 4U 1630-47 from 23-Aug-2022 to 02-Sept-2022, when the source is in high soft spectral state. The polarization properties have been analysed by three groups \cite[][]{Rawat-etal2023, Kushwaha-etal2023, Ratheesh-etal2023}, and they found that PD is an energy dependent which increases significantly  with energy. In 2-8 keV band PD is $\sim$ 0.08 and PA in the sky plane is $\sim$ 18$\degree$. They argue that the observations can explain in thin accretion disk with mildly relativistic outflowing medium.

  In the present scenario, the observed flux is described almost by Model 1a with $\langle N_{sc}\rangle$ = 1.1 (see left upper panel of Figure \ref{fig:mod1}). Therefore PD is energy independent and PD = 0.08 for $i$ = 75$\degree$.  PA is always around 90$\degree$, i.e., 
  the electric vector is parallel to the (x-y) plane or accretion disk. The energy dependency of PD can be arose by fresh scattering in optically thin ($\langle N_{sc}\rangle$ = 1)  medium (possibly, wind) where large fraction of high energy photons experience  scattering in comparison to the low energy photons, which will increase the PD value to 0.1 for high energy photons  (see table \ref{tab-gen-res} and text).

\item[]{\bf Cyg X-2:} \cite{Farinelli-etal2023} measure the polarization properties of Cyg X-2 ( a Z-source), in 2 - 8 keV band the PD = 0.018 and PA = 140$\degree$ where the polarization is in the direction of radio jet (possibly on the sky plane).
  They argue that the observed PD cannot be explained with accretion disk geometry and suggest another geometry related to the neutron star surface. The observed flux of Cyg X-2 can be described nearly with Model 1a and $\langle N_{sc}\rangle$=26.7. Thus for the inclination angle $i$ $\approx$ 60$\degree$, PD $\approx$ 0.015 and PA $\approx$ 120$\degree$, here we remind that PA is measured in a perpendicular plane to the escaped photon direction. Hence, our calculation indicates that the observed PD can be explained, in general, with accretion disk + corona geometry.

The similar range of PD ($\approx$ 0.017) is measured for atoll source {\bf GX 9+9:} by \cite{Chatterjee-etal2023}. The reported PA is $\sim$ 63$\degree$ where the range of $i$ is 40$ -$ 60$\degree$. The observed flux can be described here by Model 1a with $\langle N_{sc}\rangle$ lies in range 5 - 26. In our calculations, like Cyg X-2, the observed PD of GX 9+9 can be described in the accretion disk scenario. The calculated PA is around either (90 $-$ 120$\degree$) or (60$-$90$\degree$) for a given range of $i$ (here, 180$\degree$ differences in PA is due to the two possible definitions for  modulation curve, see equation \ref{mod-polz-scat0}).

\item[]{\bf XTE J1701-462:} {\it IXPE} had observed XTE J1701-462 during an outburst two times, Sept-2022 (epoch 1) and Oct-2022 (epoch 2). \cite{Jayasurya-etal2023} measure a significant PD $\sim$ 0.045 for epoch 1 and a negligible PD $< $ 0.01 for epoch 2 in 2 - 8 keV band. The PA is $\sim$ 143$\degree$, while the source has $i$ close to 70$\degree$.
  The observed flux can be described with Model 1a with  $\langle N_{sc}\rangle$ $>$ 26 for epoch 1 while $\sim$ 25 for epoch 2 (see Figure \ref{fig:xte} 
    for epoch 1 modeled flux). In present study with  simplistic spherical geometry we can not explain the observed PD for epoch 1, one needs other geometry. However if we exclude some fraction of forward directed escaped photons in this geometry then we can explain the observed magnitude, see left panel of Figure \ref{fig:mod1-2geo}. Moreover the observed PD for epoch 2 can be explained in present study for $i \sim$ 70$\degree$.
  The calculated PA for $i \sim$ 70$\degree$ is around 150$\degree$. Therefore, in present observations, the polarization properties of XTE J1701-462 indicates that the corona geometry changes in month scale.
For a spectro-polarimatric study in details we consider this source, see appendix \ref{sp-pol}.

\item[]{\bf Cyg X-1:} \cite{Krawczynski-etal2023} 
  estimate the polarization properties for hard state of Cyg X-1, the reported magnitude is PD $\approx$ 0.04 and PA $\approx$ 20$\degree$ in the plane of sky. They also find that the X-ray polarization almost aligns with radio jet in the plane of sky, which is consistent with the previous results of \cite{Chauvin-etal2018} using the PoGO+ balloon-borne polarimeter in 19 - 181 keV. \cite{Chauvin-etal2018} had measured PD $\approx$ 0.045 and PA = 154$\degree$ in 19 - 181 keV. The observed flux can be described by Model 2 with $\langle N_{sc}\rangle$ $\approx$ 5, thus the estimated PD $<$ 0.025. \cite{Krawczynski-etal2023} found the PD $<$ 0.03 for a spherical lamppost corona and non-spinning black hole. Therefore, our calculations are consistent with the result of \cite{Krawczynski-etal2023}. Although, 
    \cite{Krawczynski-etal2023} argue that the observed high PD can be explained with sandwich corona with $i$ $\sim$ 45$\degree$ (see theirs Figure 3).

    Clearly, like epoch 1 of  XTE J1701-462, in present study with having spherical corona we can not explain the observed high PD.  
    However,
 if we exclude the some fraction of forward escaped photons in spherical geometry then we can explain the observed high PD in broad band 2 - 181 keV (see right panel of Figure \ref{fig:mod1-2geo}).

\end{itemize}

\section{Summary and Conclusions}\label{sec:sum}
  
The polarization measurement provides two independent variables, the degree of
polarization PD and the angle of polarization PA. We observe a linearly
polarized
high energy emission (X-ray) in XRBs, AGNs. The observed polarization along
with spectra, and time variability may remove the existing degeneracy among
theoretical models mainly in terms of the radiative process and the
geometry of the emission region.
We explore the linear/ plane polarization properties in the Comptonization process using a Monte
Carlo scheme with  spherical corona geometry.
We revisit the theory of polarization in the Compton scattering process with unpolarized electrons.
We argue that the $(k \times k')$-coordinate (in which, $(k \times k')$ acts as a $z$-axis) is more suitable to describe the polarization formalism.
The single scattered unpolarized (incident) photon is polarized in perpendicular to the scattering plane. In Thomson regime it is completely polarized for scattering angle $\theta$ = 90$\degree$, see equation (\ref{deg-polz-un}) for PD as a function of $\theta$. The PD and PA can also be extracted for a given $\theta$ from its %the
modulation curve, see equation (\ref{mod_un}).
The cross section for completely polarized incident photons of polarization angle $\phi$ (or, $\theta_e$), expressed by equation (\ref{eq:cros_pol}), is not an exact expression (see the Figure \ref{fig:sigm-pol} for deviation and validation of the approximation). 
The completely polarized low-energy incident photon with %polarization angle
$\phi$ = 0 and 90$\degree$ retains its own polarization properties after scattering,
the PD as a function of $\theta$ after single scattering is expressed by equation (\ref{deg-polz-polA}) for a given $\phi$.
The Stokes parameters are invariant under Lorentz transformations, particularly in Compton scattering we argue that the value of PD and PA do not change after transforming one frame to any Lorentz-boosted frame.

For $\theta$ = 0, we find that the modulation curve of scattered photon exhibits same polarization
characteristic of incident photon, it is also valid for the range of $\theta$ $\equiv$ [0,25$\degree$]. We use this property to estimate the PD and PA of scattered photons with average scattering number
$\langle N_{sc}\rangle$, by computing the modulation curve of scattered photons of  %
$\langle N_{sc} + 1\rangle$ scattering
at $\theta = 0$.
We also compute directly the PD and PA of $\langle N_{sc}\rangle$th scattered photons 
  by using the
 pdf (corresponding to the cross section at $\theta = 0$, see equation \ref{gen-modu-p}) for known $\theta_e$ angle of $\langle N_{sc}+1\rangle$th scattering. 
Interestingly, we find that for a fixed incident photon ($k$) direction and $\theta$ $\equiv$ [0,25$\degree$] the distribution of projection of $k'_e$ (electric vector of scattered photon) on perpendicular plane to the $k$ 
also reveals the polarization properties of incident photon, which can be a working principle for Compton scattering based detector of polarization. %
We write an expression (equation \ref{avg_dpa}) for the resultant PD and PA of photons in particular direction which are a mixture of $n$th type of photons of different PD and PA.

\cite{Chandrasekhar1946} has solved the radiative transfer equations, governed
by scattering opacity by free electrons, for the two states of polarization one is along  the meridian plane and another one is normal to it. It is found that  
the emergent photons %after single scattering
from a given meridian plane are
polarized normal to the meridian plane and PD varies zero (at $i$ = 0) to 0.11 (at $i$ =90$\degree$, $i$: disk inclination angle). We obtain these results using the MC scheme in a step wise way after single scattering by discussing four relevant cases for unpolarized incident photons.
We derive these results with considering a semi spherical corona in which seed photons source situated at the centre. The different steps involved are, case I: the incident photons are along the $z$-axis (i.e., $\theta_i$ = 0) and the scattering plane is fixed to the considered emergent meridian plane; case III: the incident photon direction is fixed, say ($\theta_i$,$\phi_i$) and all possible scattering planes are considered;
case IV: the incident photon lies on the surface of cone of opening angle $\theta_i$ with having all possible scattering planes; and last a general case, Case V. 
In particular the case IV may be relevant for the external Comptonization in blazars of radio jet of opening angle $\theta_i$ \cite[][]{Kumar-Kushwaha2021}, e.g., here for $\theta_i$ = 15$\degree$ the maximum emergent photons escape with $\theta_s$ $\sim$25$\degree$ and have PD $\sim$ 0.06.

In Thomson regime for the unpolarized incident photons, the maximum PD in general case V (or law of darkening of \cite{Chandrasekhar1946}) is $\sim$ 0.11. However for a fixed scattering plane and isotropic directions of $k$, the PD is $\sim$0.33 
(see case I with $\theta_i \equiv [0,\pi]$, also right panel of Figure \ref{fig:sigm-unpol}).
In case of multi scattering we notice that after scattering number $>$ 4, the maximum PD reduces to 0.02-0.035 for all discussed cases. Here, it is noted that this multi scattered photon is basically an unpolarized photon if these photons scatter freshly with optically thin medium $\langle N_{sc}\rangle \sim 1$ then the PD  as a function of $i$ is again described by table \ref{tab-gen-res}.

We explore the energy dependency of polarization for unpolarized incident photons  by considering mainly two
different spectral sets of parameters corresponding to the hard (larger electron medium temperature, $kT_e$ = 100keV) and soft ($kT_e$ = 2.5keV) states.  
For calculations we take a simple spherical corona geometry, and estimate the
polarization of scattered photons with two average scattering numbers $\langle N_{sc}\rangle$ $\sim$ 1.1 and 5 for each spectral set. We compute for four inclination angles of disk $i$ = $\theta_s$ = 30$\degree$, 45$\degree$, 60$\degree$ and 75$\degree$, here we
have considered the ($x,y$)-plane as an accretion disk.
With spherical corona geometry, we find that the PD is independent of seed photon source temperature $kT_b$, and for $\langle N_{sc}\rangle$ $\sim$ 1.1 the PD is independent of energy in 2 - 8 keV band for both spectral sets.
Since, in Thomson regime the Comptonized flux is independent of $\theta$, thus for the unpolarized incident photons and non-relativistic corona temperature, the PD of the scattered photons would be independent of E (see appendix \ref{sp-pol}) atleast after single scattering.
The magnitude of PD as a function of $i$ is slightly lower than the values listed in table \ref{tab-gen-res} for $\langle N_{sc}\rangle$ $\sim$ 1.1 as expected.
For $\langle N_{sc}\rangle$ $\sim$ 5, the maximum value of PD is $\sim$0.03 at $i = 45\degree$.
In present convention, PA = $90\degree$ means that the electric vector is parallel to the accretion disk (or $(x,y)$-plane) for all $i$. We find that for both spectral sets with $\langle N_{sc}\rangle$ $\sim$ 1.1 the PA is $\sim$ $90\degree$ while for $\langle N_{sc}\rangle$ $\sim$ 5, the PA at $i=75\degree$ is $\sim$120$\degree$ and 140$\degree$ for soft and hard spectral sets respectively.

We also estimate the polarization for Wien spectra for both spectral sets. We
find that in Thomson regime PD has maximum value $\sim$0.03 at $i$=45$\degree$.
Since the emergent photons spectra from thin disk are a black body mainly due to the large optical depth governed by Thomson scattering (e.g., \citealp{Shakura-Sunyaev1973}, see also \citealp{Kumar-Mukhopadhyay2021}). Thus these Wien spectra calculations indicate that the polarization of emergent photons from the thin disk will
not be described by table \ref{tab-gen-res}, but it has maximum value
$\sim$0.03 for $i$=45$\degree$, see also Figure \ref{fig:mod12_wien}.

Recently, IXPE has observed many XRBs and AGNs sources and the estimated PD
for few sources is larger than 0.03, which can not be explained with considered simple spherical geometry. To understand the geometry dependency for polarization within this, %considered geometry,
we adhocly exclude the some range of the scattering angle. For a soft spectral set we obtain PD 
$\sim$ 0.045 at $i = 45\degree$ with  $\theta$ range [30,180$\degree$], and so we qualitatively obtain the observed PD value for source XTE J1701-462 \cite[][]{Jayasurya-etal2023}. Similarly, to explain the observed PD ($\sim$0.04) for Cyg X-1, \cite{Krawczynski-etal2023} conclude that it can not be obtained with simple spherical corona, which is consistent with our conclusions.
We qualitatively understand the estimated polarization properties for five sources observed by IXPE within spherical corona geometry and we almost align with the author's conclusions, except a few.
In future we intend to study the polarization properties with proper geometries of corona and also with implementation of general relativity formalism in MC scheme.

\section*{Acknowledgements}
We thank the referee for their comments and suggestions that have improved the presentation of the paper.
The work is partly supported by
the Dr. D.S. Kothari Post-Doctoral Fellowship (201718-PH/17-18/0013) of University Grant Commission (UGC), New Delhi.

\subsection*{Data availability} %No datasets are analysed.
In Figure \ref{fig:xte}, the data is taken from published work of \cite{Jayasurya-etal2023}.

\def\aap{A\&A}%
\def\aapr{A\&A~Rev.}%
\def\aaps{A\&AS}%
\def\aj{AJ}%
\def\actaa{Acta Astron.}%
\def\araa{ARA\&A}%
\def\apj{ApJ}%
\def\apjl{ApJ}%
\def\apjs{ApJS}%
\def\apspr{Astrophys.~Space~Phys.~Res.}%
\def\ao{Appl.~Opt.}%
\def\aplett{Astrophys.~Lett.}%
\def\apss{Ap\&SS}%
\def\azh{AZh}%
\def\bain{Bull.~Astron.~Inst.~Netherlands}%
\def\baas{BAAS}%
\def\bac{Bull. astr. Inst. Czechosl.}%
\def\caa{Chinese Astron. Astrophys.}%
\def\cjaa{Chinese J. Astron. Astrophys.}%
\def\fcp{Fund.~Cosmic~Phys.}%
\def\gafd{Geophys.\ Astrophys.\ Fluid Dyn.}
\def\gca{Geochim.~Cosmochim.~Acta}%
\def\grl{Geophys.~Res.~Lett.}%
\def\iaucirc{IAU~Circ.}%
\def\icarus{Icarus}%
\def\jcap{J. Cosmology Astropart. Phys.}%
\def\jcp{J.~Chem.~Phys.}%
\def\jfm{JFM}
\def\jgr{J.~Geophys.~Res.}%
\def\jqsrt{J.~Quant.~Spec.~Radiat.~Transf.}%
\def\jrasc{JRASC}%
\def\mnras{MNRAS}%
\def\memras{MmRAS}%
\def\memsai{Mem.~Soc.~Astron.~Italiana}%
\def\na{New A}%
\def\nar{New A Rev.}%
\def\nat{Nature}%
\def\natas{Nature Astronomy}%
\def\nphysa{Nucl.~Phys.~A}%
\def\pasa{PASA}%
\def\pasj{PASJ}%
\def\pasp{PASP}%
\def\physrep{Phys.~Rep.}%
\def\physscr{Phys.~Scr}%
\def\planss{Planet.~Space~Sci.}%
\def\pra{Phys.~Rev.~A}%
\def\prb{Phys.~Rev.~B}%
\def\prc{Phys.~Rev.~C}%
\def\prd{Phys.~Rev.~D}%
\def\pre{Phys.~Rev.~E}%
\def\prl{Phys.~Rev.~Lett.}%
\def\procspie{Proc.~SPIE}%
\def\qjras{QJRAS}%
\def\rmxaa{Rev. Mexicana Astron. Astrofis.}%
\def\sgg{Stud.\ Geoph.\ et\ Geod.}
\def\skytel{S\&T}%
\def\solphys{Sol.~Phys.}%
\def\sovast{Soviet~Ast.}%
\def\ssr{Space~Sci.~Rev.}%
\def\zap{ZAp}%
\def\memsai{Memorie della Societa Astronomica Italiana}

\bibliographystyle{mnras}
\bibliography{polz_pa}

\appendix
\section{$\langle U \rangle$ and $\langle Q \rangle $ for polarized incident photons} \label{pl-uq} %}

\begin{figure}[h!]\vspace{-0.19cm}
%%\captionsetup{font=footnotesize}
\centering
\begin{tabular}{l} %\hspace{-1.5cm}
    \includegraphics[width=0.37\textwidth]{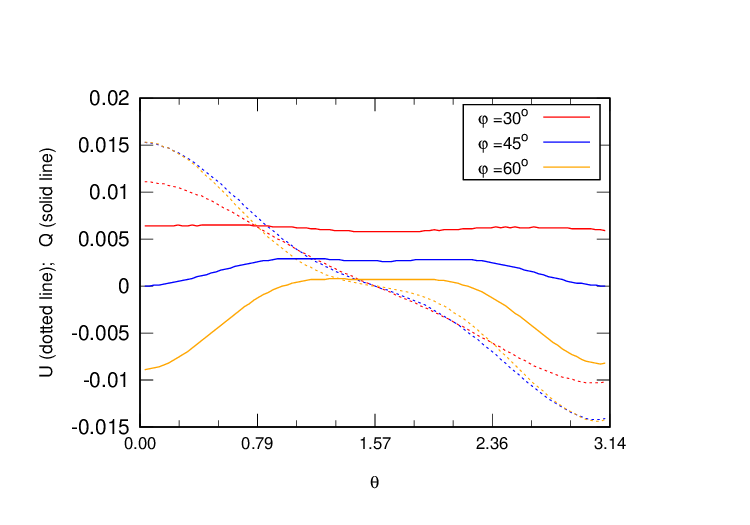} \\%
\end{tabular}\vspace{-0.3cm}
\caption{$U (\propto \sin 2\phi \cos \theta)$ and $Q (\propto (\cos^2\phi -\sin^2\phi\cos^2\theta))$ of the single scattered polarized (incident) photons as a function of $\theta$. The dotted and solid curves are for $U$ and $Q$ respectively. The red, blue and orange curves are for $\phi$ = 30, 45 and 60$\degree$ respectively.}
\label{fig:uq}
\end{figure}

Figure \ref{fig:uq} shows the Stokes parameters $U$ and $Q$ as a function of scattering angle (see section \S 2.2 for the expression) for three different values of polarization angle of  
polarized (incident) photons $\phi$ = 30, 45 and 60$\degree$. Here one can notice that for a small range of $\theta$ the $U$ dominates over $Q$. However, for all cases we find that $\langle U \rangle$ $<<$ $\langle Q \rangle$.

\section{XTE J1701-462: spectro-polarimatric comparision } \label{sp-pol}

\begin{figure}[h!]\vspace{-1.cm}
%\captionsetup{font=footnotesize}
  %\centering
\begin{tabular}{c} %\hspace{.5cm}
  \includegraphics[width=0.37\textwidth]{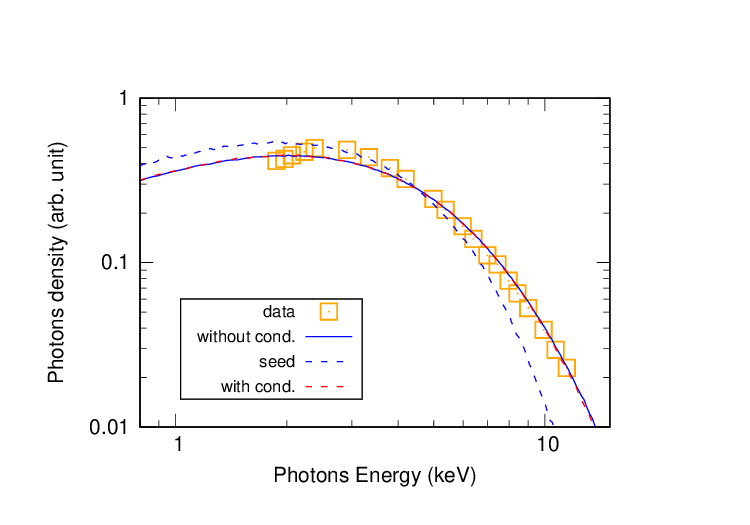} \vspace{-1.15cm}\\%sed_xte24.eps
    \includegraphics[width=0.37\textwidth]{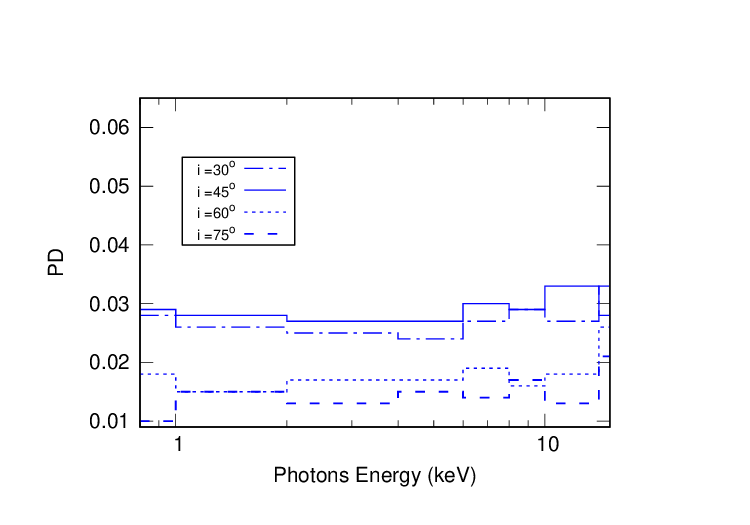} \vspace{-1.15cm}\\%xte-dp_no-cond.eps
    \includegraphics[width=0.37\textwidth]{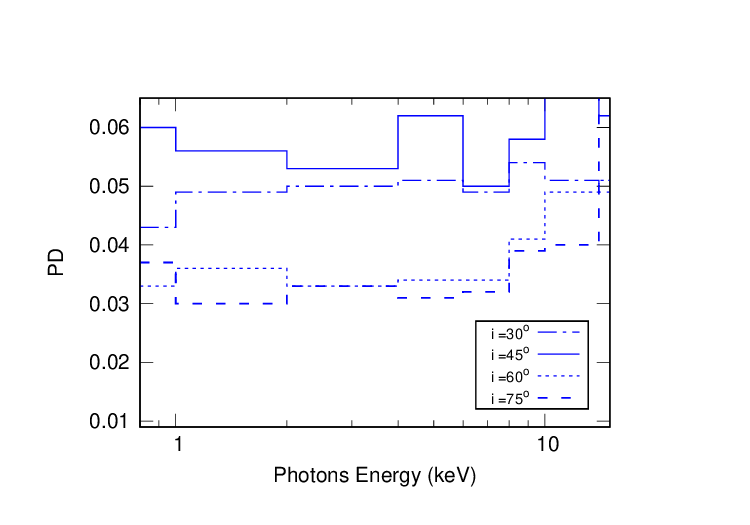} \\%xte-dp_cond.eps
\end{tabular}\vspace{-0.3cm} 
\caption{ Spectro-polarimatric measurement for source XTE J1701-462 in spherical corona. The upper panel is for the flux, here the data points are taken from \cite{Jayasurya-etal2023}, the solid curve is for Comptonized flux, the dotted curve is for seed black body flux, and the dashed curve is Comptonized flux having different $\theta$ range [45, 180$\degree$]. The parameters for Comptonization are $kT_e$= 2.5keV; $kT_b$= 1.25keV; $\langle N_{sc}\rangle$ = 24.7. The middle and lower panels are for PD with having $\theta$ range [0,180$\degree$] and [45, 180$\degree$] respectively   }
\label{fig:xte}
\end{figure}

For a observed spectro-polarimatric comparision we consider the source
  XTE J1701-462. The data points for flux are taken from \citet[][see theirs Figure 3]{Jayasurya-etal2023}. Unlike the detail modeling of \citet{Jayasurya-etal2023}, we describe the observed flux by Comptonization only, with aiming to estimate the polarization properties. The result for flux modeling in spherical geometry is shown in the upper panel of Figure \ref{fig:xte}, and the model parameters are $kT_e$= 2.5keV; $kT_b$= 1.25keV; $\langle N_{sc}\rangle$ = 24.7. We compute the PD as a function of E for two different ranges of $\theta$ $\equiv$ [0, 180$\degree$] and [45, 180$\degree$], which is shown in the middle and lower panels of Figure \ref{fig:xte} respectively.  
  We find that the variation of PA with energy is similar to the case Model Ia with  $\langle N_{sc}\rangle$ = 26.7.
  As noted earlier the observed PD of epoch 1 can not be explained in the spherical corona geometry, but one needs a different geometry where the $\theta$-angle distribution of escaped Comptonized photons does not follow the same variation as shown in Figure \ref{fig:escap-theta}.
  We also here noted that for both ranges of $\theta$ the computed fluxes are same. This is because of that in Thomson regime the scattered frequency does not depends on $\theta$ but only on angle $\alpha$ and $\alpha'$, where $\alpha$ is the angle between incident photon and incident electron, $\alpha'$ is the angle between scattered photon and incident electron, and in the lab frame it 
  is determined as
%\beqn \label{eq:sctfq_lab}
$\frac{\nu'}{\nu} = \frac{1-\frac{\text v}{c}\cos \alpha}{1-\frac{\text v}{c}\cos \alpha'+\frac{h\nu}{\gamma m_e c^2} (1-\cos \theta)}$.
%\eeqn
Hence, in general, in Thomson regime the Comptonized flux is independent of $\theta$. Further in Thomson regime and non-relativistic corona temperature, for unpolarized incident photons, the PD of the scattered photons   would be independent of E atleast after single scattering (as noted), since the PD as a function of $\theta$ is described by curve 1 of Figure \ref{fig:PD-unpolz}.

\end{document}